\documentclass[aps,prd,showpacs]{revtex4-1}
\usepackage{epsfig}
\usepackage{amsmath,amssymb,amsfonts,amsthm}
\usepackage{mathrsfs}
\usepackage{youngtab}
\usepackage[hang,nooneline]{subfigure}

\usepackage{amsmath,amssymb,amsfonts,amsthm} 
\usepackage{color}
\usepackage{graphicx}
\usepackage{pstricks,pstricks-add}
\usepackage{pst-plot}
\usepackage[hang,nooneline]{subfigure}
\usepackage{slashbox}
\usepackage{subfigure}
\definecolor{dark-green}{rgb}{0,0.7,0}
\definecolor{dark-blue}{rgb}{0,0.2,0.5}
\definecolor{med-blue}{rgb}{0,0.7,1}
\definecolor{mblue}{rgb}{0,0.2,1}
\definecolor{cnc}{rgb}{0.8,0,0}
\definecolor{light-red}{rgb}{1,0.8,0.8}
\definecolor{dark-yellow}{rgb}{1,0.8,0}
\definecolor{light-blue}{rgb}{0.8,0.9,1}
\definecolor{grey}{rgb}{0.211,0.211,0.211}
\definecolor{verylight-blue}{rgb}{0.93,0.95,1}
\definecolor{light-yellow}{rgb}{1,0.9,0.8}

\numberwithin{equation}{section}

\allowdisplaybreaks \numberwithin{equation}{section}

\newcommand{\weglassen}[1]{}

\renewcommand{\imath}{\mathrm{i}}

\begin{document}

\title{Particle motion in Ho\v{r}ava-Lifshitz black hole space-times}
\author{Victor Enolskii $^{(a),(b),(c)}$ }
\email{ V.Z.Enolskii@ma.hw.ac.uk }

\author{Betti Hartmann $^{(d)}$ }
\email{b.hartmann@jacobs-university.de}

\author{Valeria Kagramanova $^{(e)}$}
\email{va.kagramanova@uni-oldenburg.de}

\author{Jutta Kunz $^{(e)}$}
\email{jutta.kunz@uni-oldenburg.de}

\author{Claus L{\"a}mmerzahl $^{(b),(e)}$}
\email{laemmerzahl@zarm.uni-bremen.de}

\author{Parinya Sirimachan $^{(d)}$}
\email{p.sirimachan@jacobs-university.de}

\affiliation{
$(a)$ Hanse Wissenschaftskolleg (HWK), 27733 Delmenhorst, Germany\\
$(b)$ ZARM, Universit\"at Bremen, Am Fallturm, 28359 Bremen, Germany\\
$(c)$ Institute of Magnetism, 36-B. Vernadsky BLVD., Kyiv 03142, Ukraine\\
$(d)$ School of Engineering and Science, Jacobs University Bremen, 28759 Bremen, Germany\\
$(e)$ Institut f\"ur Physik, Universit\"at Oldenburg, 26111 Oldenburg, Germany}
\date\today

\begin{abstract}
We study the particle motion in the space-time of a Kehagias-Sfetsos (KS)
black hole. This is a static spherically symmetric solution of a Ho\v{r}ava-Lifshitz gravity model that 
reduces to General Relativity in the IR limit
and deviates slightly from detailed balance.
Taking the viewpoint that the model is essentially a (3+1)-dimensional modification of
General Relativity we use the geodesic equation to determine the motion of massive and massless
particles. We solve the geodesic equation exactly by using numerical techniques. We find that neither
massless nor massive particles with non-vanishing angular momentum can reach the singularity at $r=0$.
Next to bound and escape orbits that are also present in the Schwarzschild space-time we find that new types of orbits exist:
manyworld bound orbits as well as two-world escape orbits. 
We also discuss observables such
as the perihelion shift and the light deflection.
\end{abstract}
           
\maketitle

\section{Introduction}
Motivated by the study of quantum critical phase transitions Ho\v{r}ava introduced a (3+1)-dimensional quantum gravity model, later on called Ho\v{r}ava-Lifshitz (HL) gravity, 
that is power-counting renormalizable \cite{horava1,horava2} (see also \cite{sotiriou} for a recent status report).
This model reduces to General Relativity (GR) in the infrared (IR) limit, i.e. at large distances, however breaks Lorentz symmetry in the 
ultraviolet (UV), i.e.
at short distances.
The reason for this is that the model contains an anisotropic scaling with dynamical critical
exponent $z$ of the form
\begin{equation}
 \vec{r}\rightarrow b \vec{r} \ \ ,  \ \ t\rightarrow b^z t \ .
\end{equation}
In the IR the exponent becomes $z=1$ and the theory is lorentz-invariant. However, in the UV there is a strong asymmetry
between space and time with $z > 1$. In (3+1) dimensions $z=3$ \cite{horava2} and the gravity
theory becomes power-counting renormalizable. Concretely, this model breaks Lorentz invariance at short distances
because it contains only higher order spatial derivatives in the action, while higher order
temporal derivatives (which would lead to ghost degrees of freedom) do not appear. 

A number of explicit solutions of HL gravity have been found, in particular spherically symmetric
black hole solutions \cite{lu_mei_pope,ks,park,nastase}. The most general spherically symmetric solution
has been given in \cite{Kiritsis} and rotating generalizations have been studied in \cite{Ghodsi}.
One of the open problems of the model is how to couple it to matter fields. The question of how to
describe particle motion in HL gravity, i.e. to find the equivalent to the geodesic equation of GR has been addressed in 
\cite{capasso,rama,mosaffa}. In \cite{capasso} particles were studied as the optical limit of a scalar field, while
in \cite{rama} a super Hamiltonian formalism with modified dispersion relations was used. In both papers
it was found that new features arise in HL gravity such as superluminal motion and luminal motion of massive particles.
In  \cite{mosaffa} a particle action preserving foliation diffeomorphisms was introduced and it was
found that massless particles follow GR geodesics, while the trajectories of massive particles depend on their mass.
In most studies of test particle motion the hypothetical corrections to the GR geodesics were neglected 
\cite{bobo,liu_lu_yu_lu,iorio1,gwak,iorio2,iorio3,chen_wang,HKL,LoboHarko11, HGKHL,HTAA10, bobo,Konoplya}.

In this paper we take the latter viewpoint and study solutions to the GR geodesic equation in HL black hole space-times, in particular in the
space-time of a Kehagias-Sfetsos (KS) solution, a static and spherically symmetric solution to HL gravity with vanishing
cosmological constant. The geodesic motion in this space-time has been studied previously \cite{bobo,liu_lu_yu_lu,iorio1,
gwak,iorio2,iorio3,chen_wang,HKL,LoboHarko11, HGKHL,HTAA10, bobo,Konoplya} and a number of constraints on
the parameters of HL gravity have been found. Observables such as the perihelion shift and the light deflection
were also studied in these papers, however, either approximations were used or only circular orbits were studied.
In this paper we are aiming at solving the geodesic equation exactly by using numerical techniques and at exploring the
complete set of solutions of the geodesic equation.

Our paper is organized as follows: in Section II we give the model and the black hole solutions.
In Section III we give the geodesic equation, while Section IV contains our results. We conclude in Section V.

\section{The model}
\subsection{The action}

The model proposed by Ho\v{r}ava \cite{horava1,horava2} uses the ADM decomposition of the metric that reads as follows
\begin{equation}
 ds^2 = - N^2 dt^2 + g_{ij} \left(dx^i + N^i dt\right)\left(dx^j + N^j dt\right)  \ ,
\end{equation}
where $N(t,x^i)$ and $N^i(t,x^i)$ are the lapse and shift functions, respectively, and $g_{ij}(t,x^i)$ is the 
3-metric with $i,j=1,2,3$. In \cite{horava2} it was assumed that the theory is invariant under space-independent time reparametrization
and time-dependent spatial diffeomorphisms, i.e. under
\begin{equation}
 t\rightarrow \tilde{t}(t) \ \ \ , \ \ \ x^i \rightarrow \tilde{x}^i (t,x^i)   \ ,
\end{equation}
which restricts the lapse function to depend only on $t$. The action proposed in \cite{horava2} then reads
\begin{eqnarray}
\label{HLaction}
S=\tilde{S}_0 + S_0 + S_1  \ ,
\end{eqnarray}
where
\begin{eqnarray}
 \tilde{S}_0= \int dt d^3 x \sqrt{g} N \left[\frac{2}{\kappa^2} \left(K_{ij} K^{ij} - \lambda K^2\right)\right]  \ , \
S_0= \int dt d^3 x \sqrt{g} N \left[
\frac{\kappa^2 \mu^2}{8(1-3\lambda)}\left( \Lambda_{W} R 
- 3\Lambda_{W}^2\right)\right]
\end{eqnarray}
and
\begin{eqnarray}
 S_1=\int dt d^3 x \sqrt{g} N \left[\frac{\kappa^2\mu^2(1-4\lambda)}{32(1-3\lambda)} R^2 - \frac{\kappa^2}{2w^4} C_{ij}C^{ij} + \frac{\kappa^2\mu}{2w^2} \varepsilon^{ijk} R_{il} \nabla_j R^l_k - \
\frac{\kappa^2\mu^2}{8} R_{ij}R^{ij}\right] \ ,
\end{eqnarray}
where 
\begin{equation}
 K_{ij} =\frac{1}{2N} \left(\frac{\partial g_{ij}}{\partial t} - \nabla_i N_j - \nabla_j N_i\right) \ , \ 
 C^{ij}=\varepsilon^{ikl} \nabla_k \left(R^j_l - \frac{1}{4} R \delta^j_l\right)  \  .
\end{equation}
$g$ is the determinant of the metric $g_{ij}$ and $R_{ij}$, $\nabla_i$ correspond to the spatial components
of the covariant derivative and the Ricci tensor, respectively. $C^{ij}$ is the Cotton tensor
and $\lambda$, $\kappa$, $\mu$, $w$ and $\Lambda_W$ are constants. The integrand of $-(S_0+S_1)$ is interpreted
as the potential part, while the integrand of $\tilde{S}_0$ is interpreted as the kinetic
part. 

In the IR limit the action is dominated by $\tilde{S}_0+S_0$ and reduces to the Einstein-Hilbert action for
\begin{equation}
\lambda=1 \ , \ c=\frac{\kappa^2 \mu}{4} \sqrt{\frac{\Lambda_W}{1-3\lambda}} \ \ , \ \ G_{N} = \frac{\kappa^2}{32\pi c} \ \ , \ \
\Lambda=\Lambda_W \ ,
\end{equation}
where $c$ is the speed of light, $G_{N}$ is Newton's constant and $\Lambda$ is the cosmological constant. Note that for $\lambda > 1/3$, i.e.
in particular for $\lambda=1$, the
constant $\Lambda_W$ and hence $\Lambda$ should be negative.  In the following we will set $\lambda=1$ (unless otherwise stated) and consider the
additional terms of $S_1$ as a (3+1)-dimensional modification of General Relativity. The action as given above satisfies the
requirement of detailed balance which essentially means that the potential $V$ in the Ho\v{r}ava-Lifshitz action
derives from a superpotential $W$:
\begin{equation}
 V=E^{ij} {\cal G}_{ijkl} E^{kl}  \ \ , \ \ E^{ij}=\frac{1}{\sqrt{g}} \frac{\delta W}{\delta g_{ij}}  
\end{equation}
and ${\cal G}^{ijkl}=\frac{1}{2}(g^{ik}g^{jl} + g^{il} g^{jk})-\lambda g^{ij} g^{kl}$ is the DeWitt metric.
The requirement of detailed balance drastically reduces the number of invariants to consider in the potential $V$. 

The problem with the theory as stated above is that for $\lambda \approx 1$ it predicts the wrong sign of the 4-dimensional
cosmological constant. Moreover the detailed balance condition is chosen solely to simplify the theory. Hence, theories that
violate detailed balance have been considered. In \cite{ks} the following term was added to the action
\begin{equation}
\label{sv1}
 S_{v} = \int dt d^3 x \sqrt{g} N \frac{\kappa^2 \mu^2}{8(3\lambda-1)} \omega R  \ ,
\end{equation}
where $\omega$ is an arbitrary constant. In the $\Lambda_{W}=0$ limit which we are mainly interested in here
the Einstein-Hilbert action is recovered in the IR for
\begin{equation}
 \lambda=1 \ \ , \ \ G_N=\frac{\kappa^2}{32\pi c} \ \ , \ \ c^2 = \frac{\kappa^4 \mu^2}{16(3\lambda-1)}\omega  \ .
\end{equation}

\subsection{Spherically symmetric solutions}
Kehagias and Sfetsos (KS) found a spherically symmetric, static  black hole solution 
to a Ho\v{r}ava-Lifshitz gravity model
with action $S+S_v$ for $\Lambda_{\rm W}=0$ and $\lambda=1$. The Ansatz for the metric is
\begin{equation}
\label{metric}
 ds^2 = N^2(r) dt^2 - f^{-1}(r) dr^2 - r^2 \left(d\theta^2 + \sin^2\theta d\varphi^2\right) 
\end{equation}
and the solution reads
\begin{equation} 
\label{sol_ks}
 N^2=f=1+\omega r^2 - \sqrt{\omega^2 r^4 + 4\omega m r}  \  ,
\end{equation}
where $\omega= 16\mu^2/\kappa^2$ and $m$ is an integration constant. 
In \cite{iorio1} constraints on the value of $\omega m^2$ were found 
by comparing the perihelion shift in the KS space-time with observations in the solar
system. It was found that $\omega m^2 \ge 7.2\cdot 10^{-10}$ for Mercury,
 $\omega m^2 \ge 9\cdot 10^{-12}$ for Mars and $\omega m^2 \ge 1.7\cdot 10^{-12}$ for Saturn.
Moreover, a similar comparison gave $\omega m^2 \ge 8\cdot 10^{-10}$ for the S2 star 
orbiting the supermassive black hole in our galaxy as well as $\omega m^2 \ge 1.4\cdot 10^{-18}$
for extrasolar planets \cite{iorio2}. In \cite{bobo} constraints from
innermost stellar circular orbits (ISCOs) for certain black holes were considered
and it was found that $\omega \simeq 3.6 \cdot 10^{-24} cm^{-2}$ (in appropriate units).
In \cite{liu_lu_yu_lu} the light deflection
in the solar system was used to constrain the parameter. It was found that
$\omega m^2 \ge 1.17\cdot 10^{-16}$ for Earth, $\omega m^2 \ge 8.28\cdot 10^{-17}$
for Jupiter and  $\omega m^2 \ge 8.28\cdot 10^{-15}$ for the Sun.
The IR limit of (\ref{sol_ks}) is given by the Schwarzschild solution $N^2=f=1-2m/r$. The Kretschmann
scalar $K=R_{\mu\nu\rho\sigma} R^{\mu\nu\rho\sigma}$ reads
\begin{equation}
 K=\left(\frac{\partial^2 f}{\partial r^2}\right)^2 + \frac{4}{r^2} \left(\frac{\partial f}{\partial r}\right)^2
+ \frac{4 f^2}{r^4} - \frac{8f}{r^4} + \frac{4}{r^4}  \ ,
\end{equation}
which for small $r$ behaves like $1/r^3$. Hence the solution possesses a physical singularity at $r=0$ \cite{ks} and two horizons at
\begin{equation}
\label{horizon}
 r_{\pm}=m\pm \sqrt{m^2-\frac{1}{2\omega}}
\end{equation}
as long as $\omega m^2 \ge 1/2$. Note that the corrections from Ho\v{r}ava-Lifshitz gravity now allow for
the existence of up to two horizons. The extremal solution has $\omega m^2=1/2$ and $r_+=m$. 
The Hawking temperature of black hole solutions is given by $T_{\rm H}=\kappa/(2\pi)$, where $\kappa$ is the surface
gravity that for static solutions is given by
\begin{equation}
 \kappa^2=-\frac{1}{4} g^{tt} g^{ij} (\partial_i g_{tt})(\partial_j g_{tt})  \ .
\end{equation}
For the KS solution we find 
\begin{equation}
 T_{\rm H} = \frac{1}{2\pi} \frac{\omega(r_{\pm}-m)}{1+\omega r_{\pm}^2}  \ ,
\end{equation}
which in the $\omega\rightarrow \infty$ limit tends to the known Schwarzschild result $T_{\rm H}=(8\pi m)^{-1}$. 
Obviously, the extremal solutions with $r_+=m$ have $T_{\rm H} =0$. For more details about the thermodynamics
of black holes in Ho\v{r}ava-Lifshitz gravity see e.g. \cite{Cai_ohta}.

\section{Solutions to the geodesic equation in Ho\v{r}ava-Lifshitz black hole space-times}

For a general static spherically symmetric solution of the form (\ref{metric}) the Lagrangian $\mathcal{L}_{\rm g}$ 
for a point particle reads
\begin{eqnarray}
\label{lagrangian_geo}
\mathcal{L}_{\rm g}=\frac{1}{2}g_{\mu\nu}\frac{dx^{\mu}}{ds}\frac{dx^{\nu}}{ds}=\frac{1}{2}\varepsilon
=\frac{1}{2}\left[N^2\left(\frac{dt}{d\tau}\right)^{2}-\frac{1}{f}\left(\frac{dr}{d\tau}\right)^{2}-r^2
\left(\frac{d\theta}{d\tau}\right)^{2}-r^2 \sin^2\theta\left(\frac{d\varphi}{d\tau}\right)^{2}\right]
 \ ,  \end{eqnarray}
where $\varepsilon=0$ for massless particles and $\varepsilon=1$ for massive particles, respectively. 

The constants of motion are the energy $E$ and the angular
momentum (direction and absolute value) of the particle. We choose
$\theta=\pi/2$ to fix the direction of the angular momentum and have
\begin{eqnarray}
E:=N^2\frac{dt}{d\tau}\ \ , \ \
L_z:=r^2\frac{d\varphi}{d\tau}  \ \ .
\end{eqnarray}
Using these constants of motion we get
\begin{eqnarray}
\label{eq1}
 \left(\frac{dr}{d\tau}\right)^2 = \frac{f}{N^2}\left(E^2 - \tilde{V}_{\rm eff}(r) \right) 
\end{eqnarray}
and
\begin{eqnarray}
\label{eq2}
\left(\frac{dr}{d\varphi}\right)^2 = \frac{r^4}{L_z^2} \frac{f}{N^2}\left(E^2 - \tilde{V}_{\rm eff}(r) \right)  \ ,
\end{eqnarray}
where $\tilde{V}_{\rm eff}(r)$ is the effective potential
\begin{equation}
\label{potential}
 \tilde{V}_{\rm eff}(r)=N^2\left( \varepsilon + \frac{L_z^2}{r^2}    \right)  \ .
\end{equation}
In the following we will consider the KS black hole solution (\ref{sol_ks}). 
The geodesic equation (\ref{eq2}) then becomes 
\begin{eqnarray}
\label{geo1}
 \left(\frac{1}{r}\frac{dr}{d\varphi}\right)^4 + 2\left(\frac{1}{r}\frac{dr}{d\varphi}\right)^2  
P(r) = Q(r)  \ ,
\end{eqnarray}
where
\begin{equation}
 P(r)=\frac{1}{L_z^2}\left(\omega \varepsilon r^4 + (\varepsilon-E^2+\omega L_z^2)r^2  + L_z^2\right)
\end{equation}
and
\begin{eqnarray}
 Q(r)&=&\frac{1}{L_z^4}\left[-
2\varepsilon \omega (E^2-\varepsilon)r^6 
-4\omega m \varepsilon^2 r^5 
+ (-2 \omega E^2 L_z^2 +  4 \omega L_z^2 \varepsilon  +(E^2-\varepsilon)^2) r^4  \right. \nonumber \\
&-& \left. 8\omega m L_z^2 \varepsilon r^3 + 2L_z^2 (\omega L_z^2 - E^2 + \varepsilon) r^2  -4\omega m L_z^4  r 
+ L_z^4\right] \ .
\end{eqnarray}
For massive particles ($\varepsilon=1$) the order of the polynomials $P(r)$ and $Q(r)$ is $4$ and $6$, respectively, 
while for massless
particles ($\varepsilon=0$) it is $2$ and $4$. 

Rewriting (\ref{geo1}) we find
\begin{equation}
\label{sol_geo1}
 \varphi-\varphi_0 = \pm \int\limits_{r_0}^{r} \frac{dr}{r\sqrt{-P \pm \sqrt{P^2 + Q}}} \ .
\end{equation}

The motion of test particles
in KS black hole space-times has been studied extensively before \cite{bobo,liu_lu_yu_lu,iorio1,gwak,iorio2,iorio3,chen_wang}, 
however, it was not attempted to find the complete set of solutions. This is what
we are aiming at here. 
The integral on the right hand of (\ref{sol_geo1}) cannot be solved in terms of hyperelliptic functions, at least not to our knowledge.
However, an analytic treatment seems possible in some limiting cases. This will be reported elsewhere \cite{ehkkls_new}. In this paper 
we solve the geodesic equation (\ref{geo1})
numerically.

\section{Results}

\subsection{The effective potential}
In order to understand which types of orbits are possible in the KS space-time, we first study the effective
potential. To make contact with the Schwarzschild case we rewrite (\ref{eq1})
as follows
\begin{equation}
 \left(\frac{dr}{d\tau}\right)^2 ={\cal E} - V_{\rm eff}(r)  \ ,
\end{equation}
where ${\cal E}=E^2-\varepsilon$ and 
\begin{equation}
\label{veff}
 V_{\rm eff}(r)= 
\tilde{V}_{\rm eff}(r)-\varepsilon=\left(\omega r^2 - \sqrt{\omega^2 r^4 + 4\omega m r}\right)\left(\varepsilon + \frac{L_z^2}{r^2}\right)
+ \frac{L_z^2}{r^2}  \ . 
\end{equation}
For $ r \gg (4m/\omega)^{1/3}$ this effective potential becomes
$V_{\rm eff}(r\gg (4m/\omega)^{1/3})\approx -2m\varepsilon/r - 2mL_z^2/r^3 + L_z^2/r^2$, which is just the
effective potential in the Schwarzschild space-time.

\begin{figure}[h!]
  \begin{center}
    \subfigure[$\omega = 0.50$, $\varepsilon = 1$]{\label{fig1a}\includegraphics[scale=0.30]{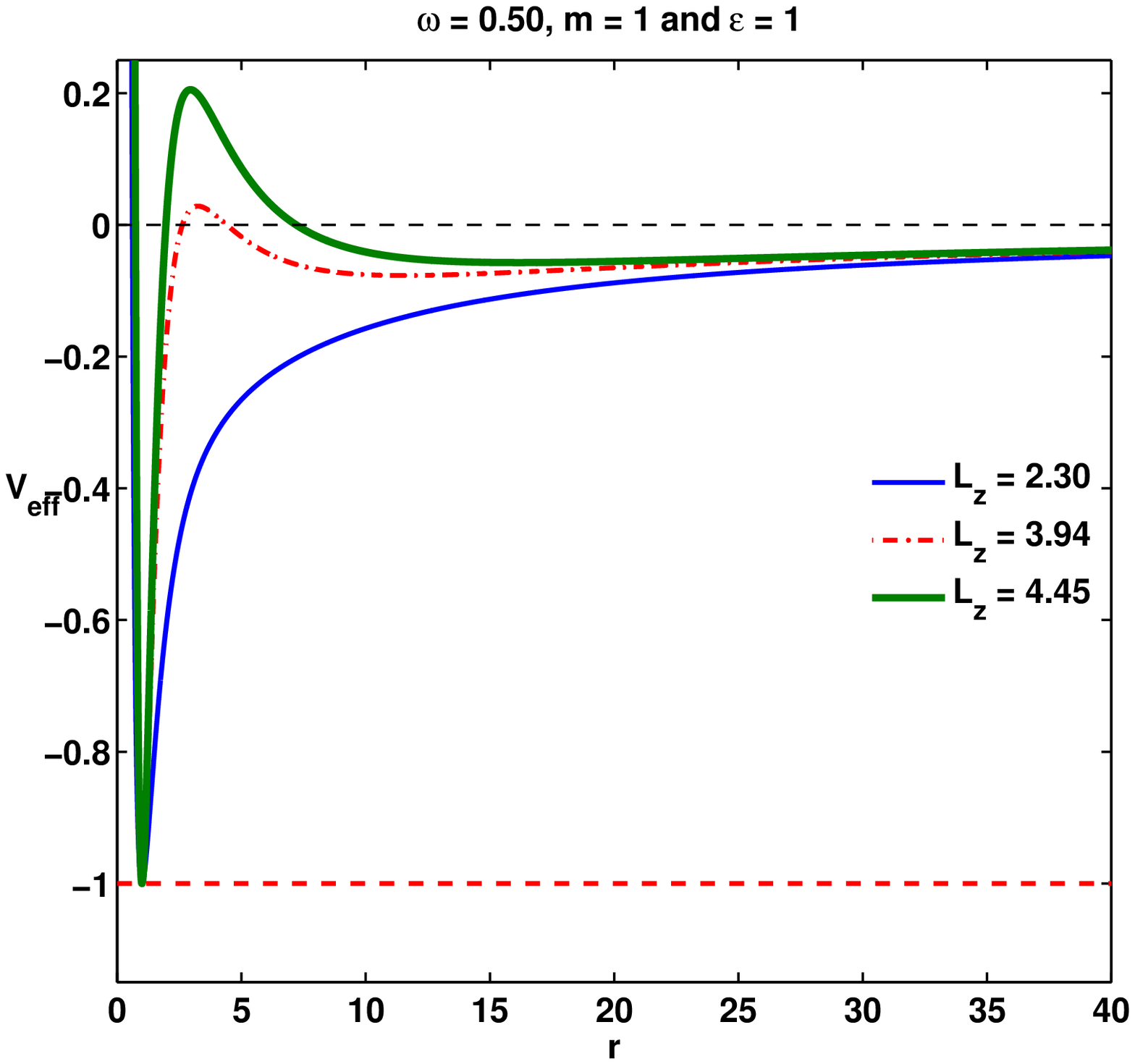}}
    \subfigure[$\omega = 0.52$, $\varepsilon = 1$]{\label{fig1b}\includegraphics[scale=0.30]{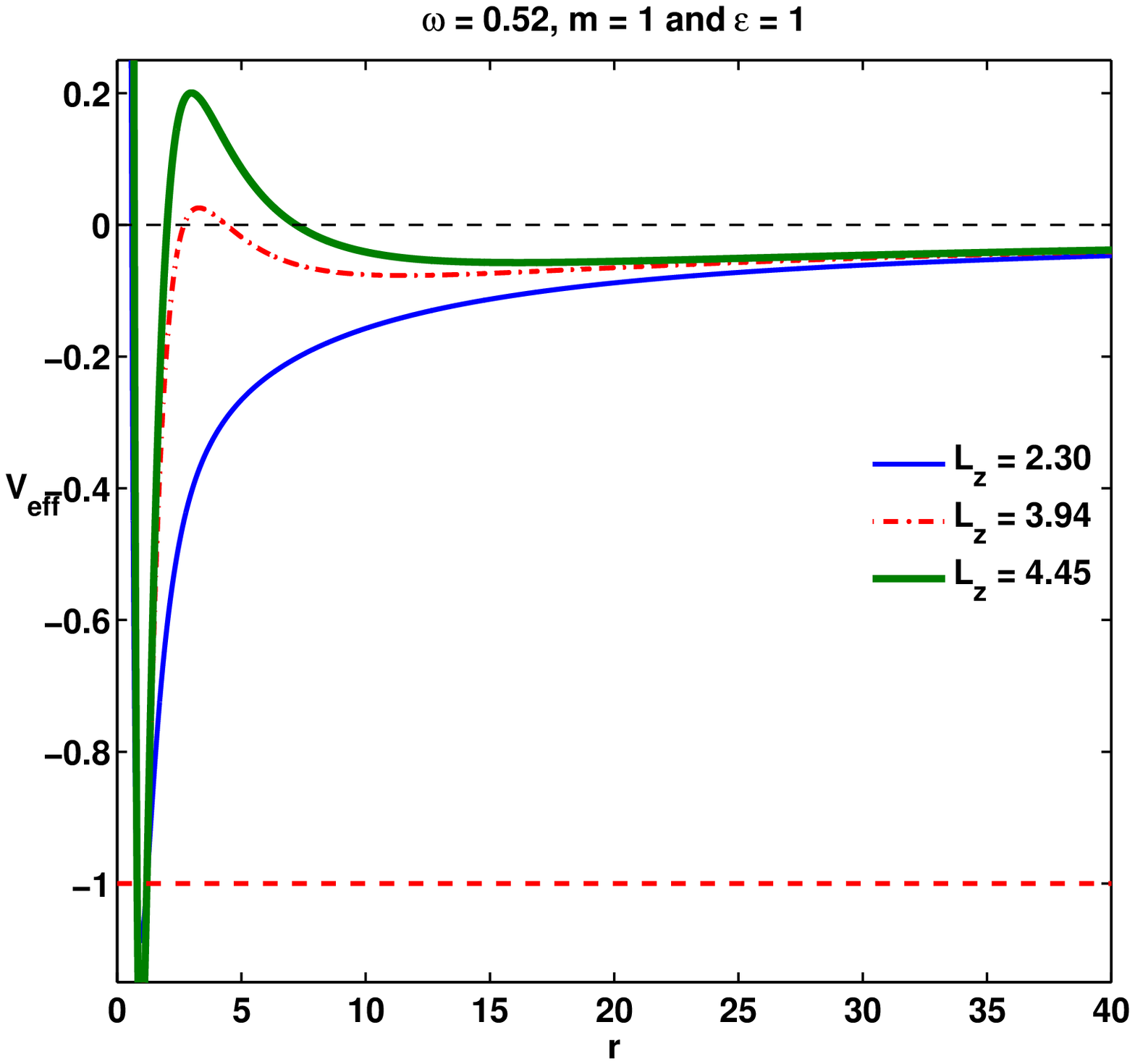}}
    \subfigure[$\omega = 5.1$ $\times$ $10^4$, $\varepsilon = 1$]{\label{fig1c}\includegraphics[scale=0.30]{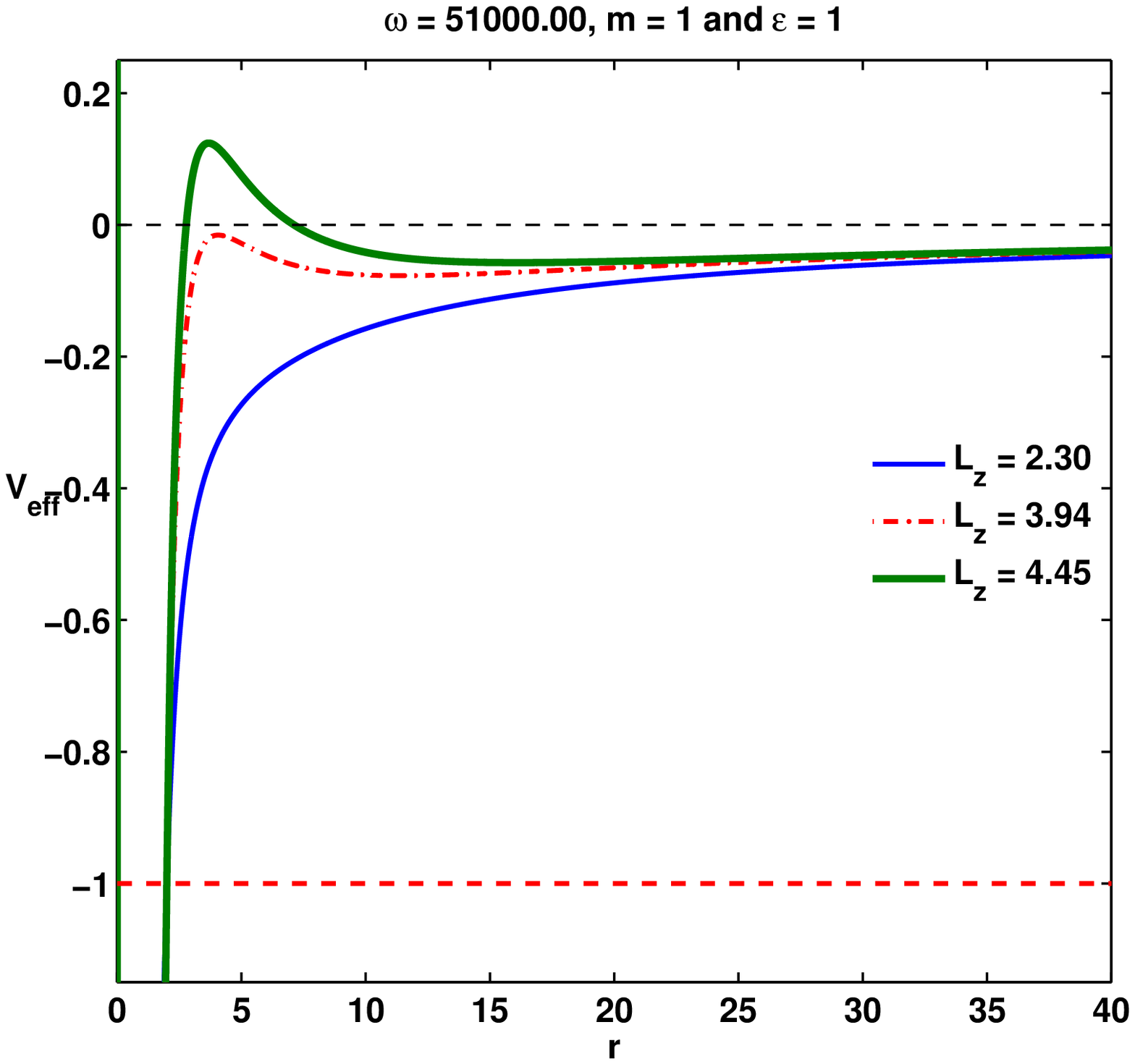}}\\
    \subfigure[$\omega = 0.50$, $\varepsilon = 0$]{\label{fig1d}\includegraphics[scale=0.30]{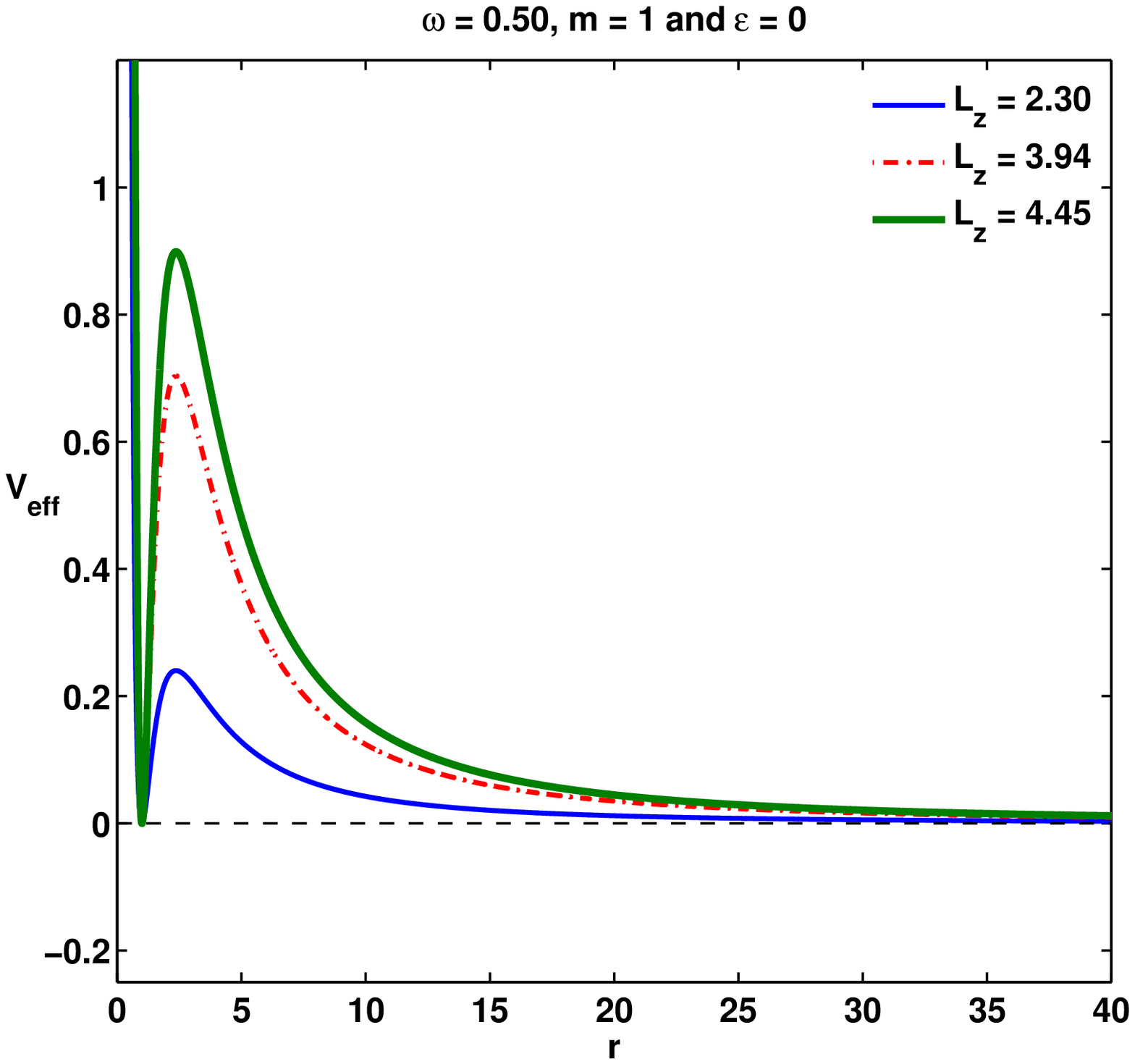}}
    \subfigure[$\omega = 0.52$, $\varepsilon = 0$]{\label{fig1e}\includegraphics[scale=0.30]{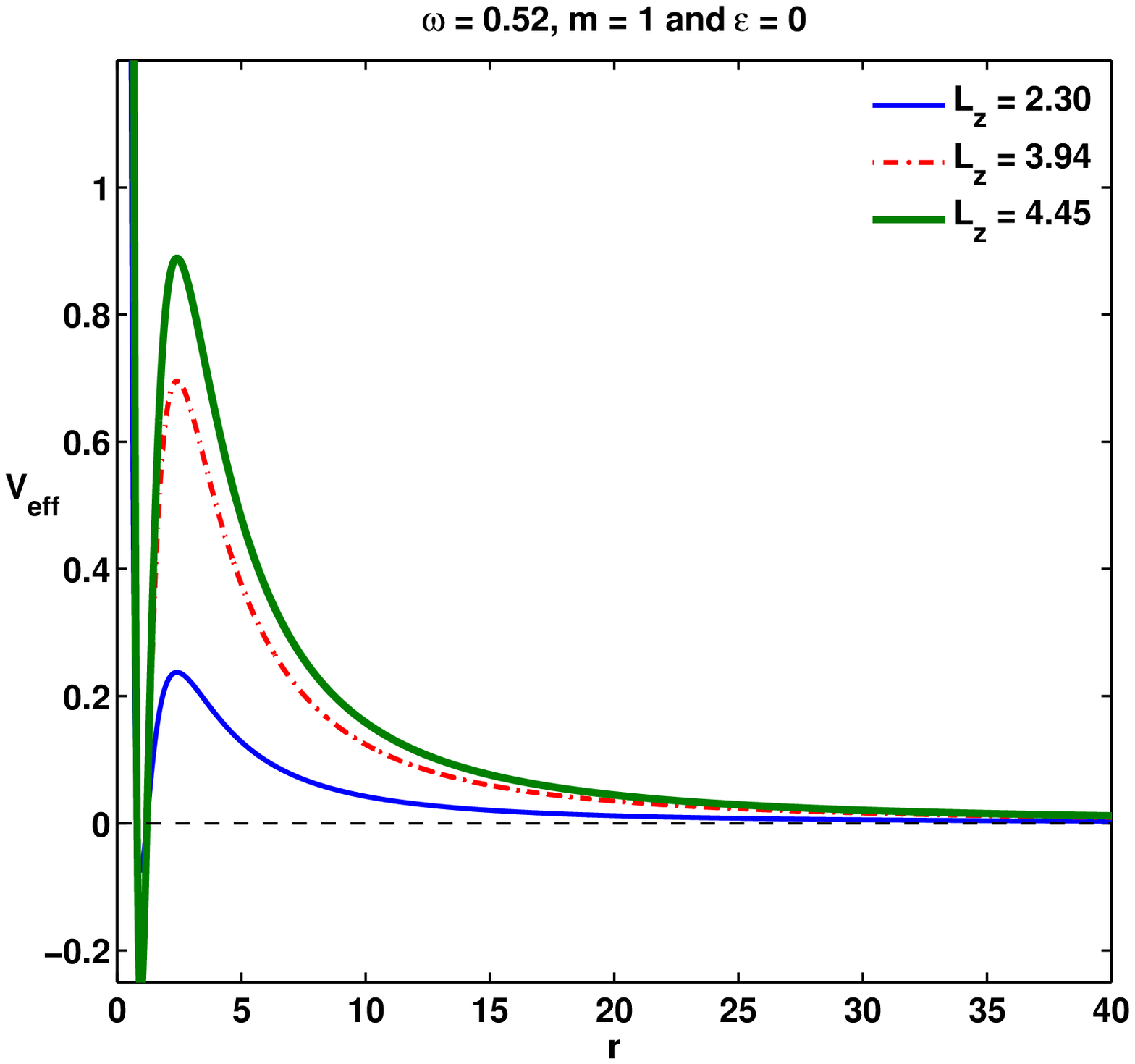}}
    \subfigure[$\omega = 5.1$ $\times$ $10^4$, $\varepsilon$ = 0]{\label{fig1f}\includegraphics[scale=0.30]{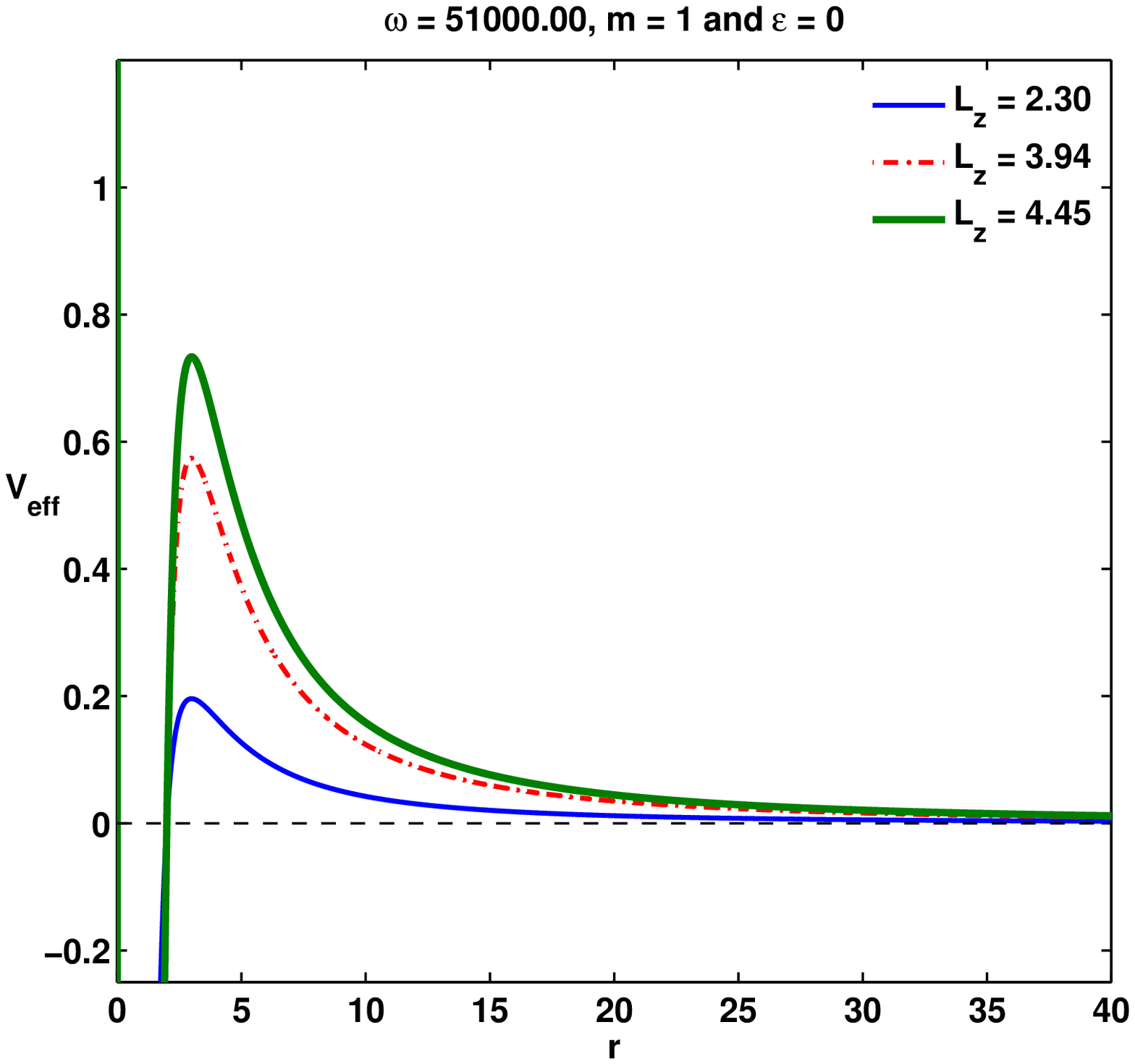}}

    \end{center}
   \caption{The effective potential $V_{\rm eff}(r)$ for a massive ((a)-(c)) and a massless ((d)-(f)) test particle, respectively, 
for different values of $\omega$ and $L_z$.}
  \end{figure}

The first point to note is that while for $\omega\rightarrow\infty$ the potential at $r\ll 1$ behaves
like $V_{\rm eff}(r \ll 1) \approx -2mL_z^2/r^3$ (this is just the Schwarzschild limit), it behaves
like $V_{\rm eff}(r\ll 1) \approx L_z^2/r^2$ for generic $\omega$. Hence there is a positive infinite angular
momentum barrier for both massive and massless test particles which does not exist in the Schwarzschild limit.
The first conclusion is hence that test particles with non-vanishing angular momentum cannot reach the
singularity at $r=0$ in the KS space-time. Moreover, for the extremal solution with $r=r_+=m$ we find that
$dV_{\rm eff}(r)/dr\vert_{r=r_+}=0$ and $V_{\rm eff}(r=r_+)=-\varepsilon$.

\begin{figure}[h!]
  \begin{center}
   \includegraphics[scale=0.40]{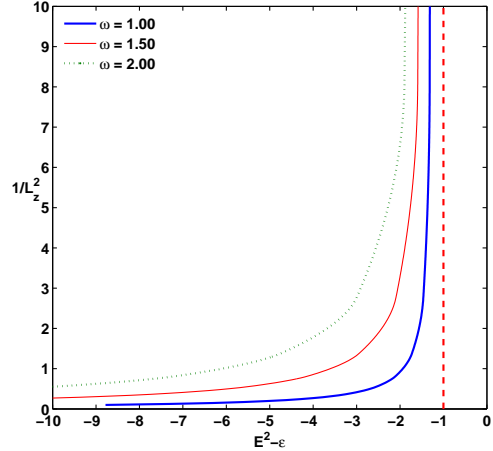}
  \end{center}
\caption{The values of $E^2-\varepsilon$ and $1/L_z^2$ corresponding to the absolute minimum of the effective
potential $V_{\rm eff}(r)$ at small $r$ for different values of $\omega$, $m=1$ and $\varepsilon=1$. Note that while here we treat $E^2-\varepsilon$ 
as a parameter that can take arbitrary values, we should have 
$E^2-\varepsilon\geq -1$ when looking for zeros of ${\cal E}-V_{\rm eff}(r)$.}
\label{fig2a}
\end{figure}

\begin{figure}[h!]
  \begin{center}
  \includegraphics[scale=0.45]{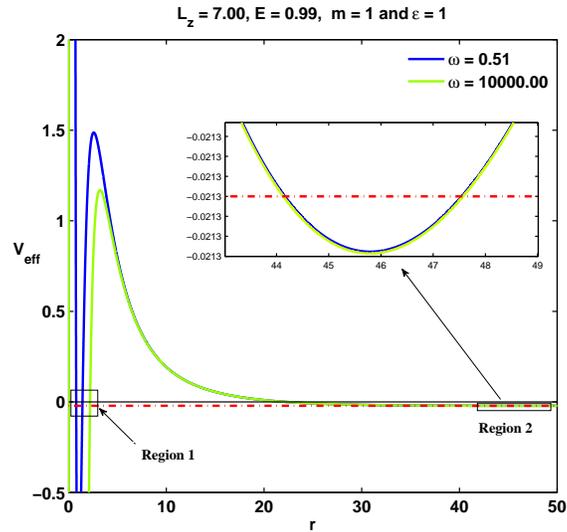}
    \end{center}
  \caption{The two regions of the potential for which bound orbits of massive test particles exist. In region 1, we have
manyworld bound orbits, while in region 2 there exist bound orbits. Here $L_z = 7.0$, $E^2 = 0.9787$,
$m = 1.0$, while $\omega=0.51$  and $\omega=10^4$, respectively. 
The red dotted-dashed line represents the total energy ($E^2$-$\varepsilon$) of the test particle. } 
\label{bound_orbits}
  \end{figure}

\begin{figure}[h!]
  \begin{center}
    \subfigure[$\omega=5.1$]{\label{fig2b}\includegraphics[scale=0.4]{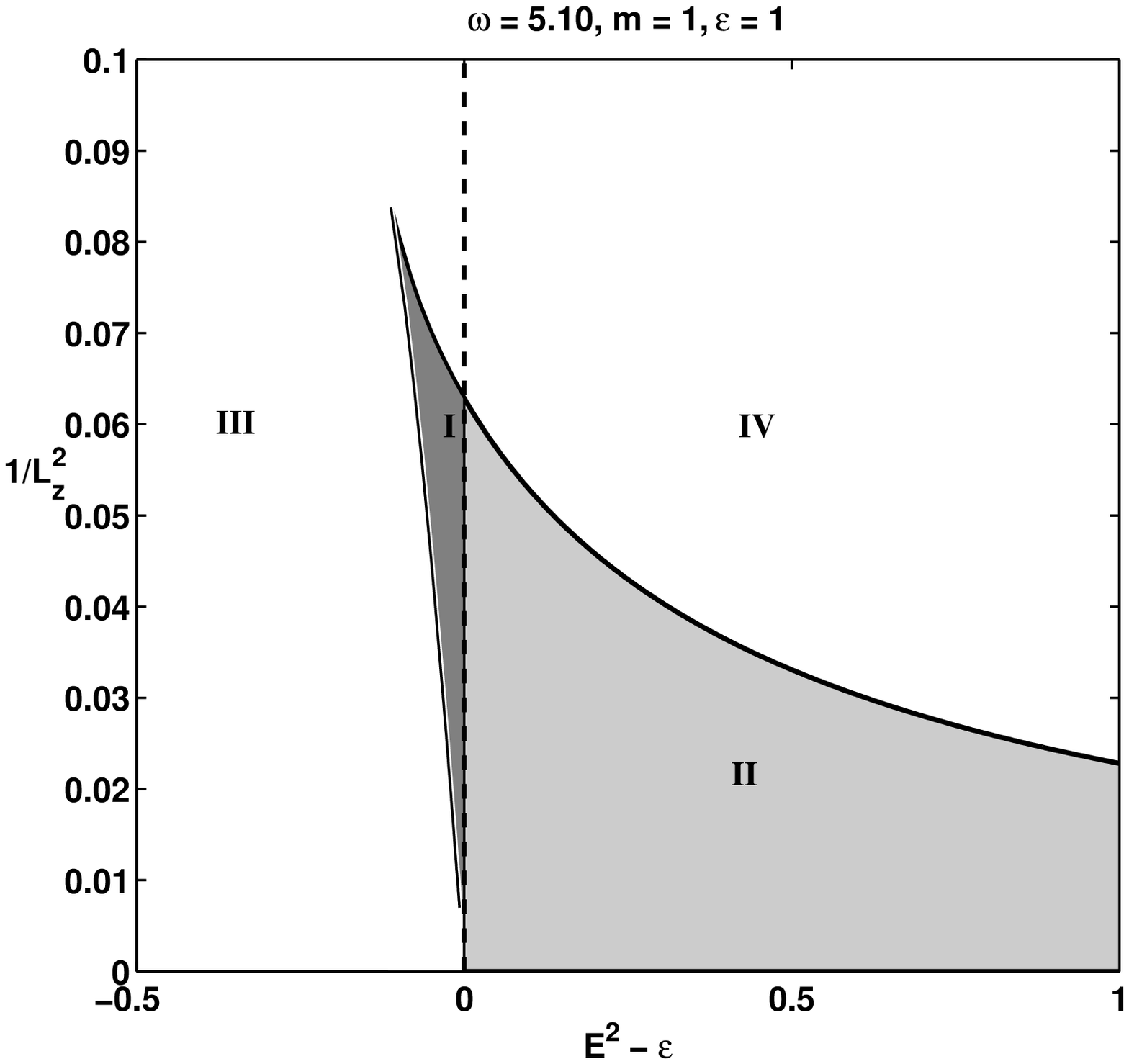}}
    \subfigure[$\omega=5.1\cdot 10^4$]{\label{fig2c}\includegraphics[scale=0.4]{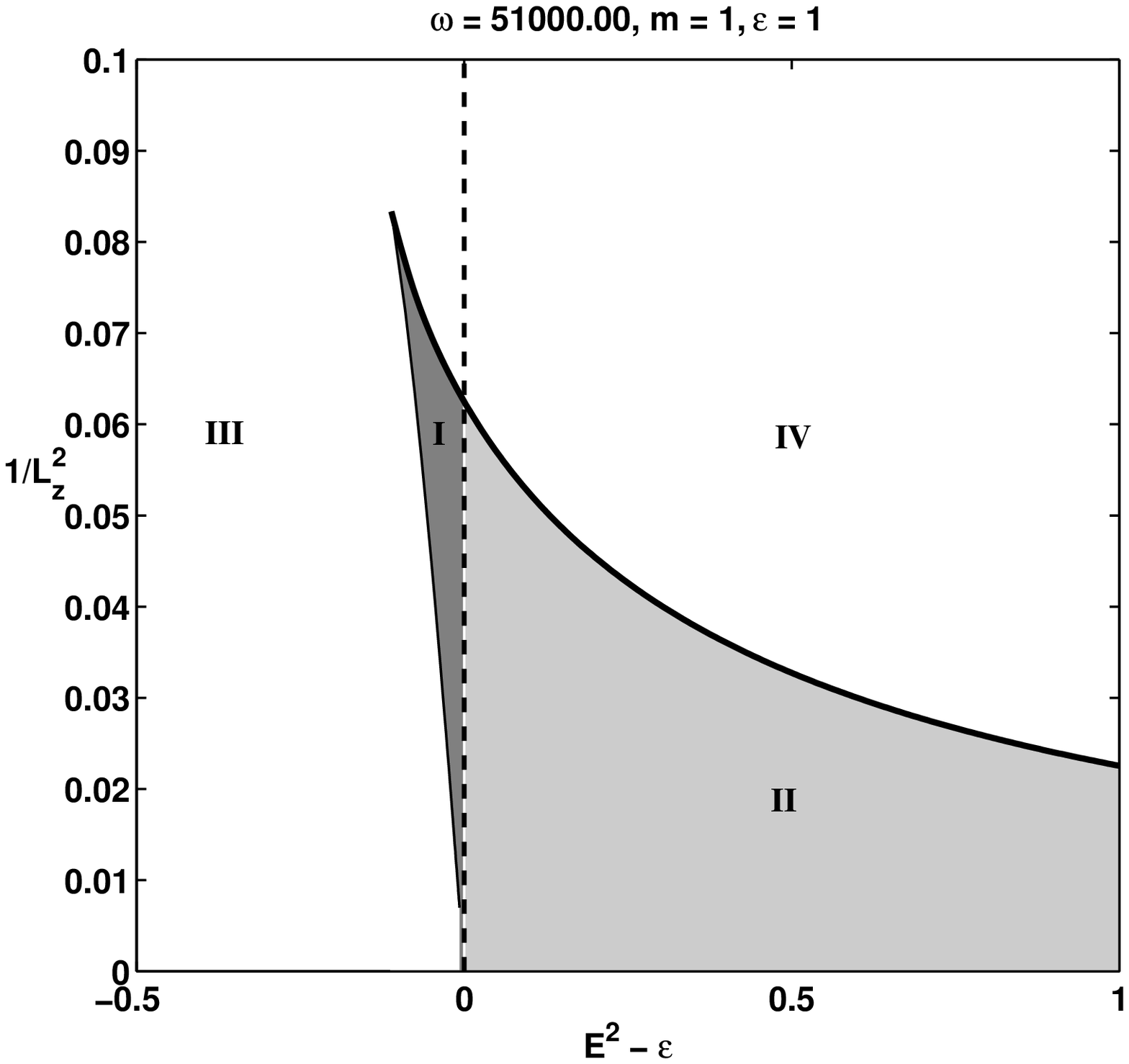}}\\
    \end{center}
\caption{The values of $E^2-\varepsilon$ and $1/L_z^2$ corresponding 
to the  maximum (thick upper line) and relative minimum at large $r$ (thin lower
line) of the effective potential $V_{\rm eff}(r)$ are given for
$m=1$, $\varepsilon=1$, $\omega=5.1$ (left) and $\omega=5.1 \cdot 10^4$ (right), respectively. In the dark shaded
region (region I) there exist manyworld bound orbits (MBO) and bound orbits (BO), while in the light shaded region 
(region II) manyworld bound orbits (MBO)
as well as escape orbits (EO) exist. In region III there are manyworld bound orbits (MBO), while there are 
two-world escape orbits (TEO) in region IV.}
\end{figure}

On the other hand, for particles without angular momentum $L_z=0$, the effective potential is always negative and behaves like
$V_{\rm eff}(r \ll 1)\approx -\varepsilon \sqrt{4\omega m r}$ for small $r$, while it is equivalent to the Schwarzschild
potential for large $r$: $V_{\rm eff}(r \gg 1)\approx -2m\varepsilon/r$. 


\subsubsection{Massive test particles}

In Figs. \ref{fig1a}-\ref{fig1c} we show how the effective potential $V_{\rm eff}(r)$ for a massive test particle ($\varepsilon=1$)
changes for different values of $L_z$ and $\omega$ and $m=1$.

It is obvious that the effective potential at large $r$ doesn't change much when decreasing $\omega$ from the
Schwarzschild limit $\omega=\infty$. Hence, the types of orbits available for large $r$ are very similar
to the Schwarzschild case. This is not surprising since Ho\v{r}ava-Lifshitz gravity is a gravity theory
that is supposed to modify General Relativity at short distances, but has no effects
on the long distance physics. 
In comparison to the Schwarzschild case, the
effective potential possesses a further minimum at small $r$. This is represented by the
curves in Fig.\ref{fig2a}. In this latter plot, we assume that $E^2-\varepsilon$ is a parameter that can have all possible
values to show that an additional minimum exists, but keep in mind that to find the zeros of 
${\cal E}-V_{\rm eff}(r)$ we need to require ${\cal E} \geq -1$.
Note that the value of this minimum
is negative and always smaller than $-1$. It increases for 
decreasing $\omega$ and becomes equal to $-1$ in the extremal limit. This is clearly seen in Figs.\ref{fig1a}-\ref{fig1c}.

\begin{figure}[h!]
  \begin{center}
  \includegraphics[scale=0.45]{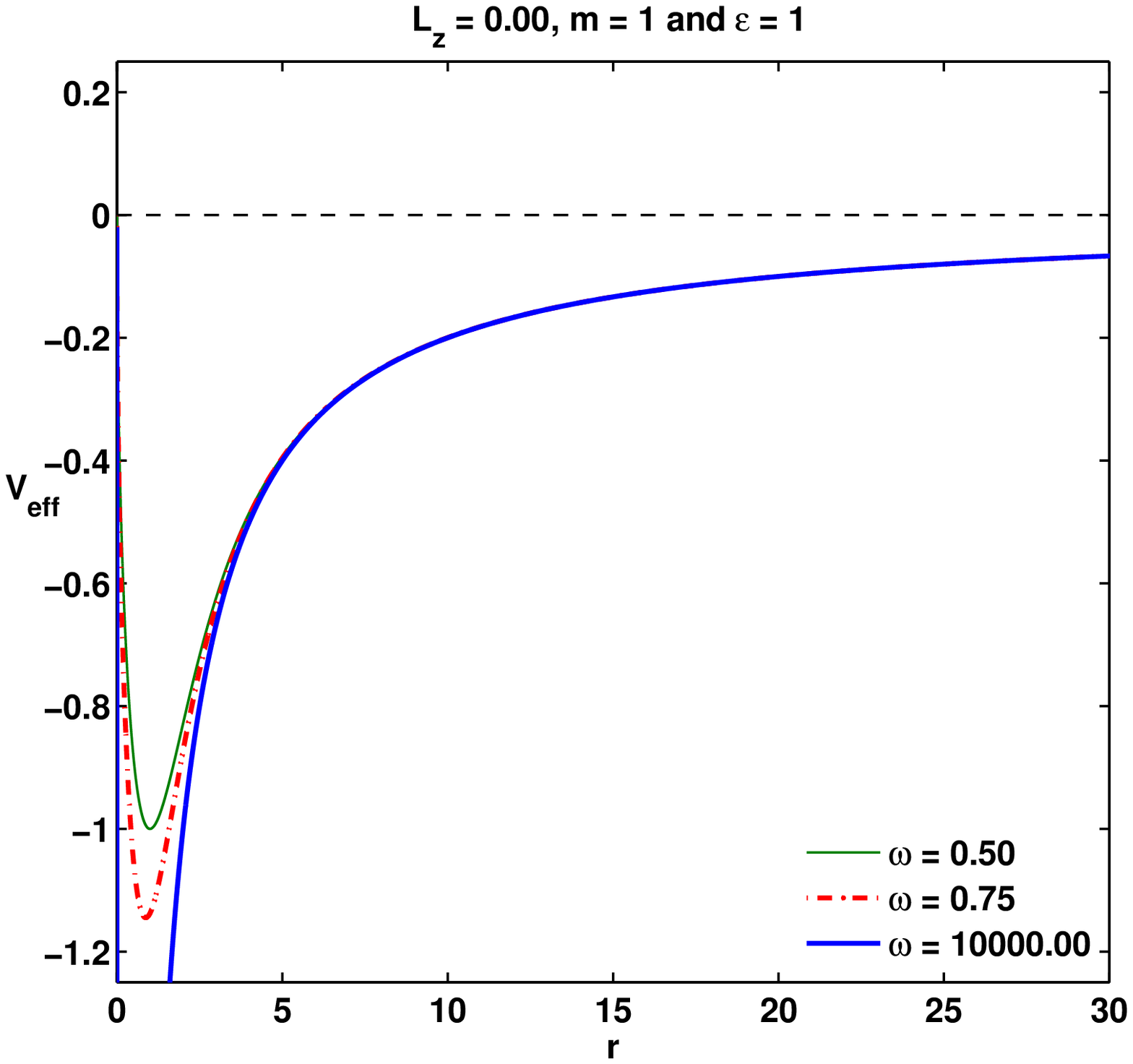}
    \end{center}
 \caption{The effective potential $V_{\rm eff}(r)$ for radial trajectories ($L_z=0$) of massive particles ($\varepsilon=1$)
in the space-time of a KS black hole with $m=1$ and different values of $\omega$. }
\label{radial_potential}
  \end{figure}

\begin{figure}[h!]
  \begin{center}
    \subfigure[Region 1, $\omega = 0.51$]{\includegraphics[scale=0.3]{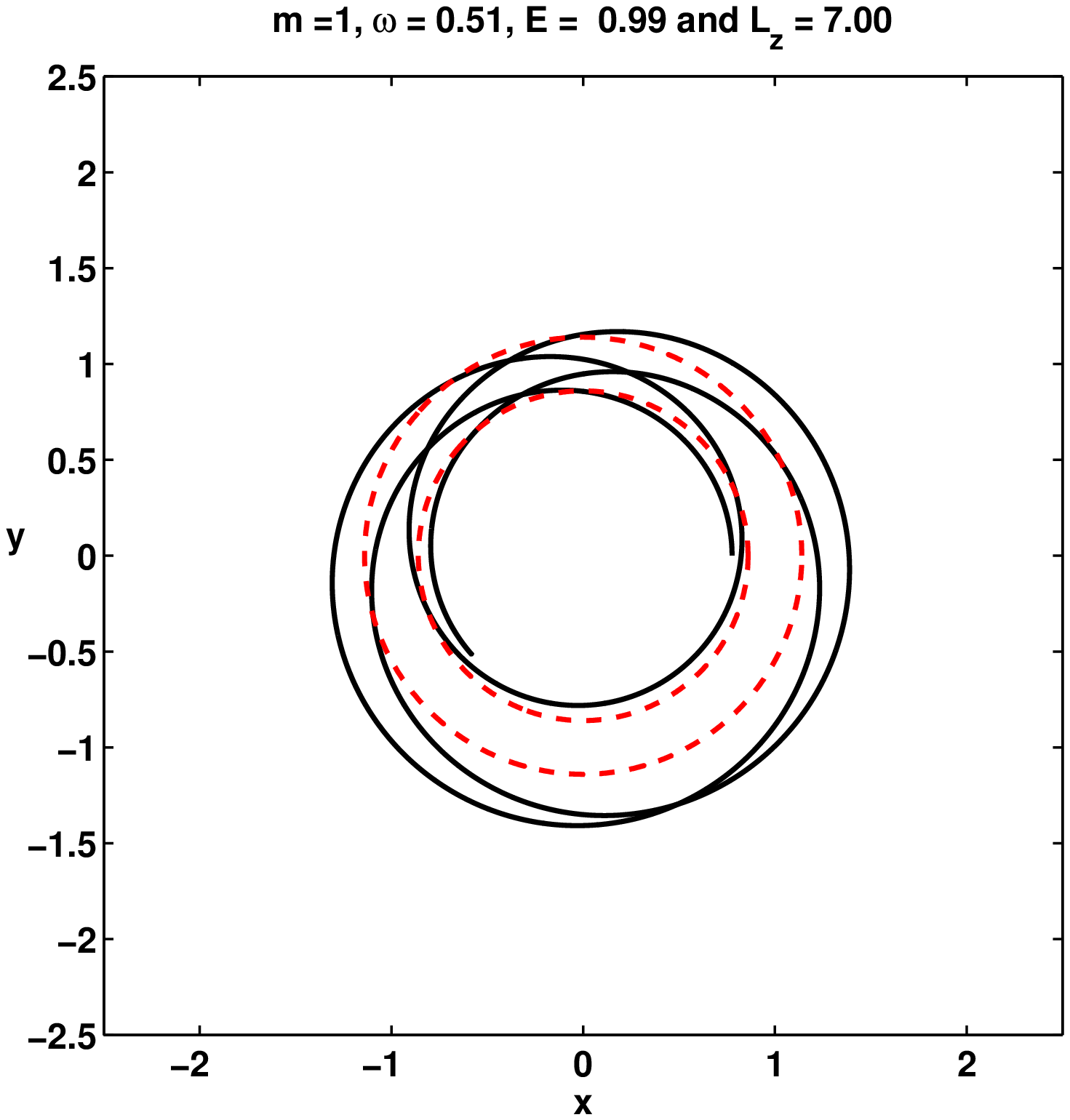}}
    \subfigure[Region 1, $\omega = 10^4$]{\includegraphics[scale=0.3]{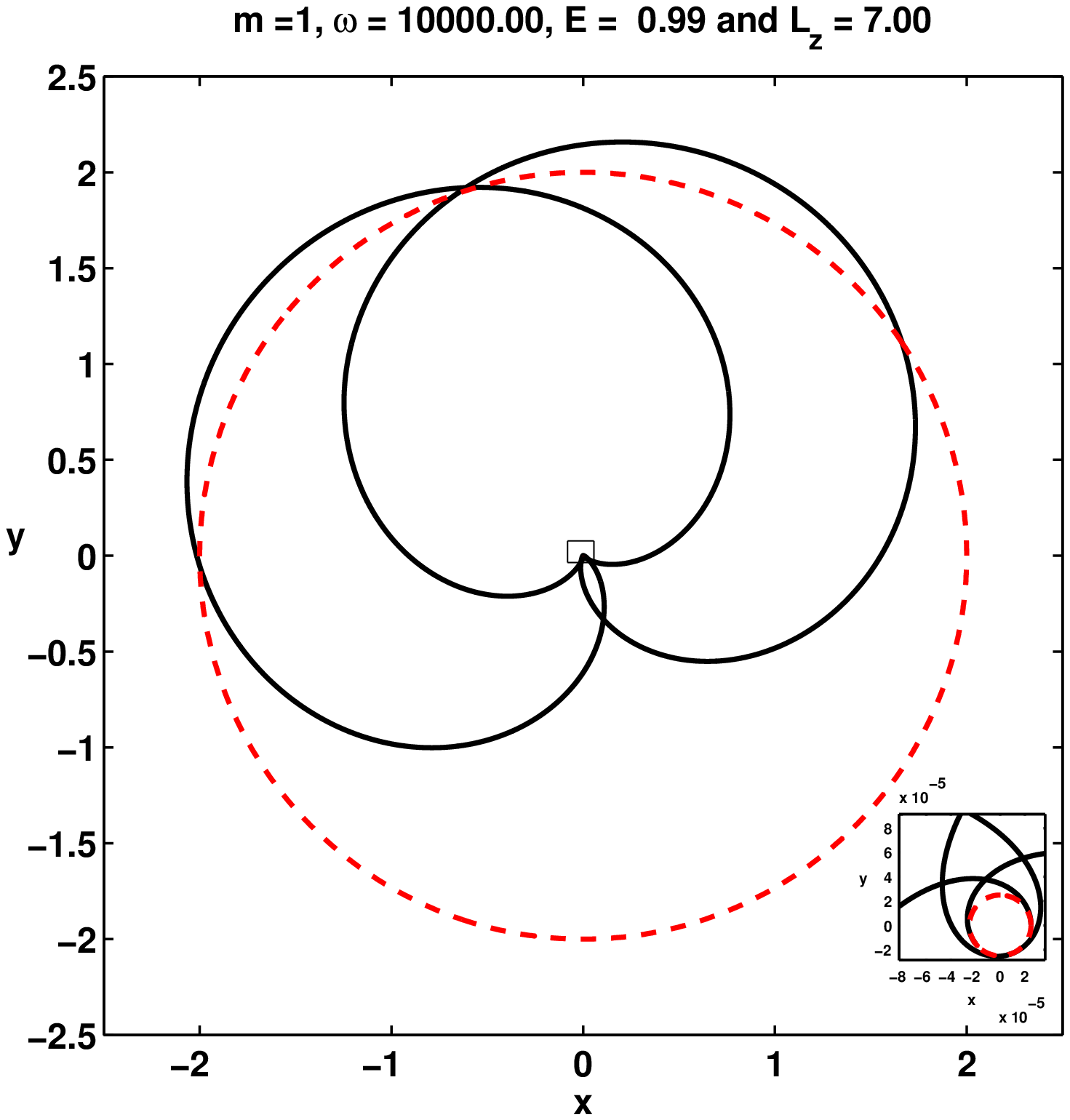}}\\
    \subfigure[Region 2, $\omega = 0.51$]{\label{fig3c}\includegraphics[scale=0.3]{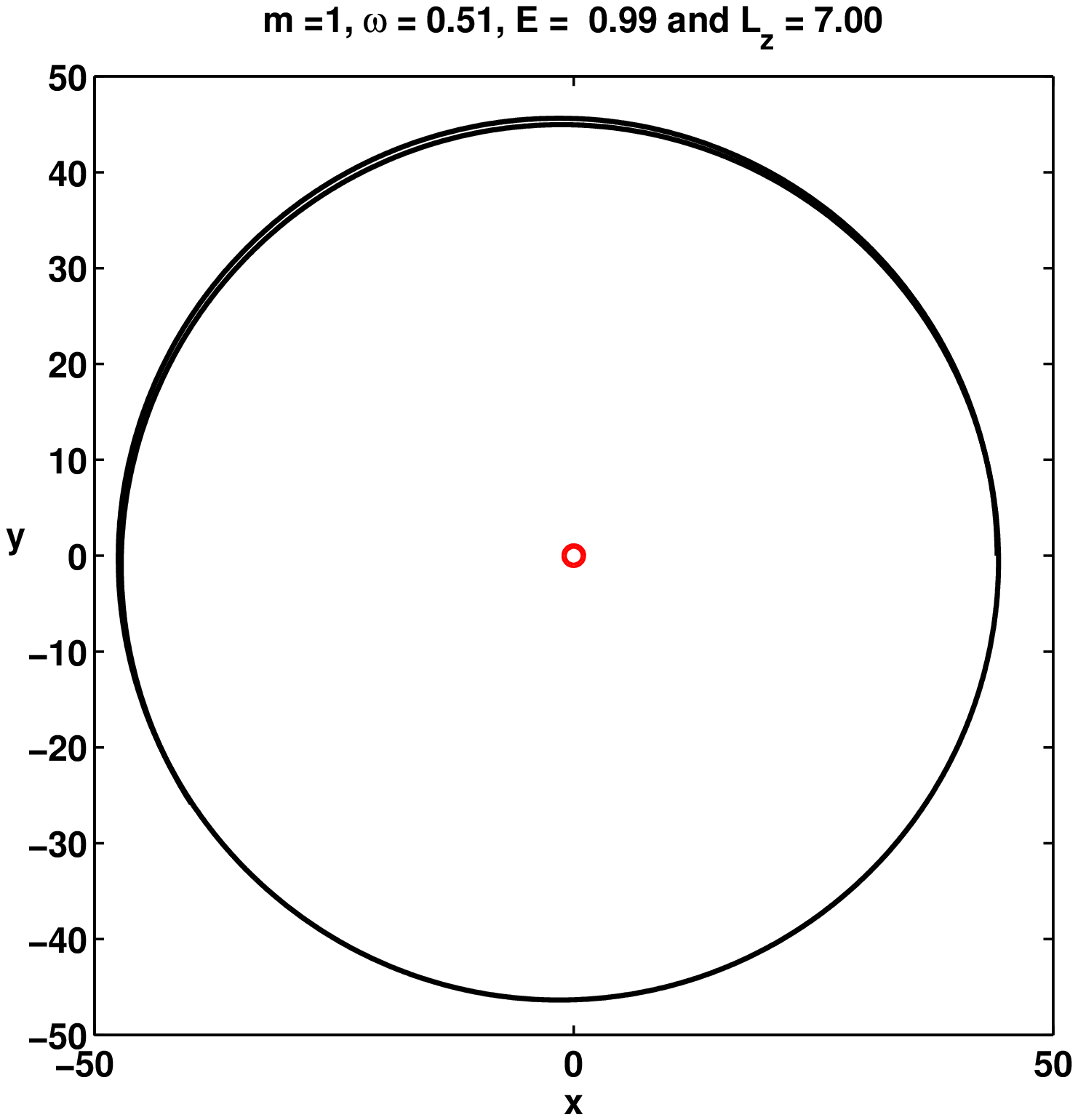}}
    \subfigure[Region 2, $\omega = 10^4$]{\label{fig3d}\includegraphics[scale=0.3]{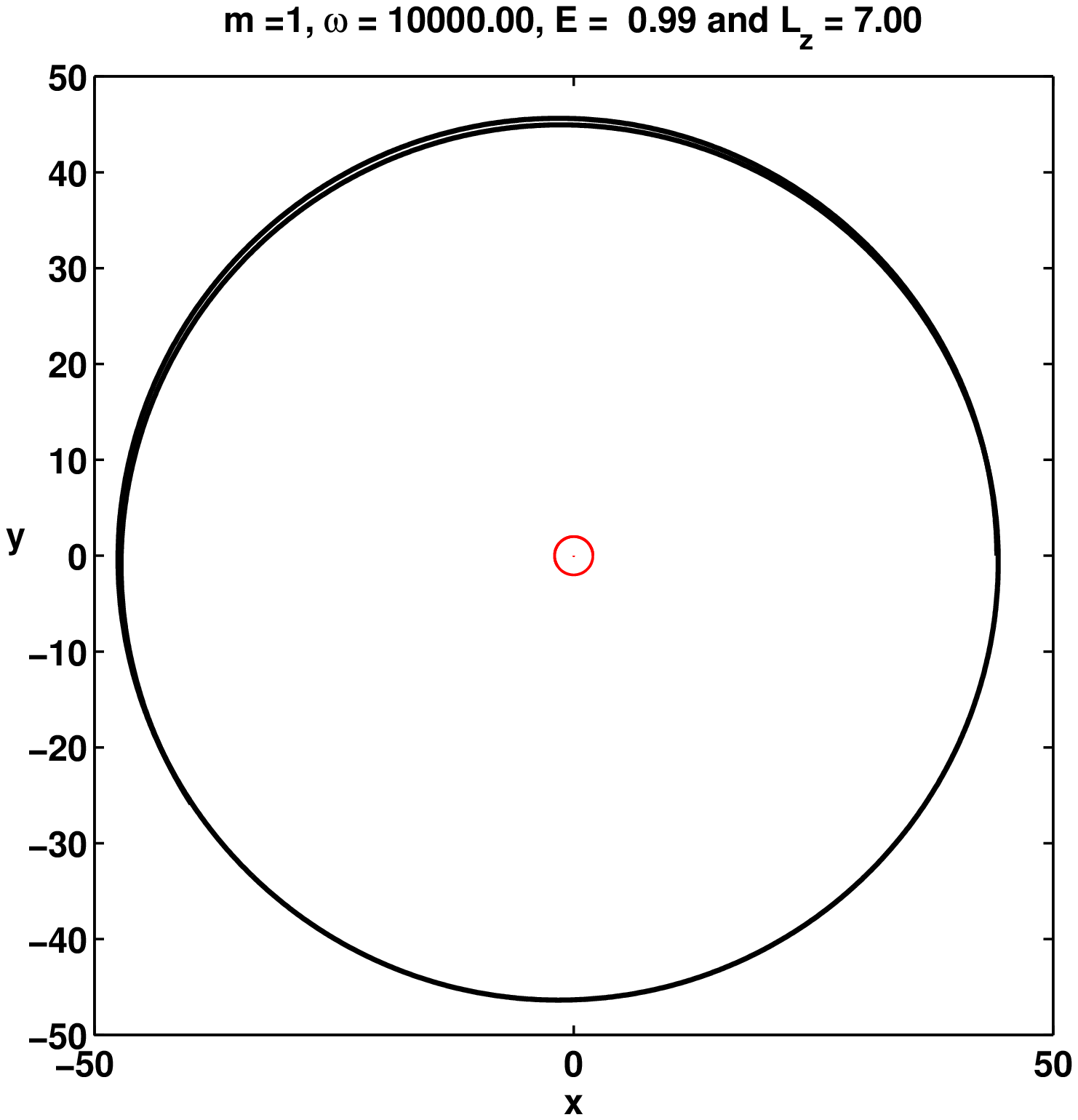}}   
    \end{center}
   \caption{Examples of manyworld bound orbits (MBO) and bound orbits (BO) of a massive test particle ($\varepsilon = 1$) with $L_z = 7.00$, $E^2 = 0.9787$ 
in the space-time of a KS black hole with $m = 1.00$ as well as $\omega=0.51$ (left) and $\omega= 10^4$ (right).
We show manyworld bound orbits (MBO) (region 1, top)  and  bound orbits (BO) (region 2, bottom), respectively. The red dashed circles in the 
plot represent the horizons of the KS black hole. Note that we are plotting two radial periods during which
the particle moves from $r_{\rm min}$ to $r_{\rm max}$ and back again.}
\label{fig3}  
\end{figure}

\begin{figure}[h!]
  \begin{center}
    \subfigure[$\omega=0.51$]{\label{orbit2a}\includegraphics[scale=0.30]{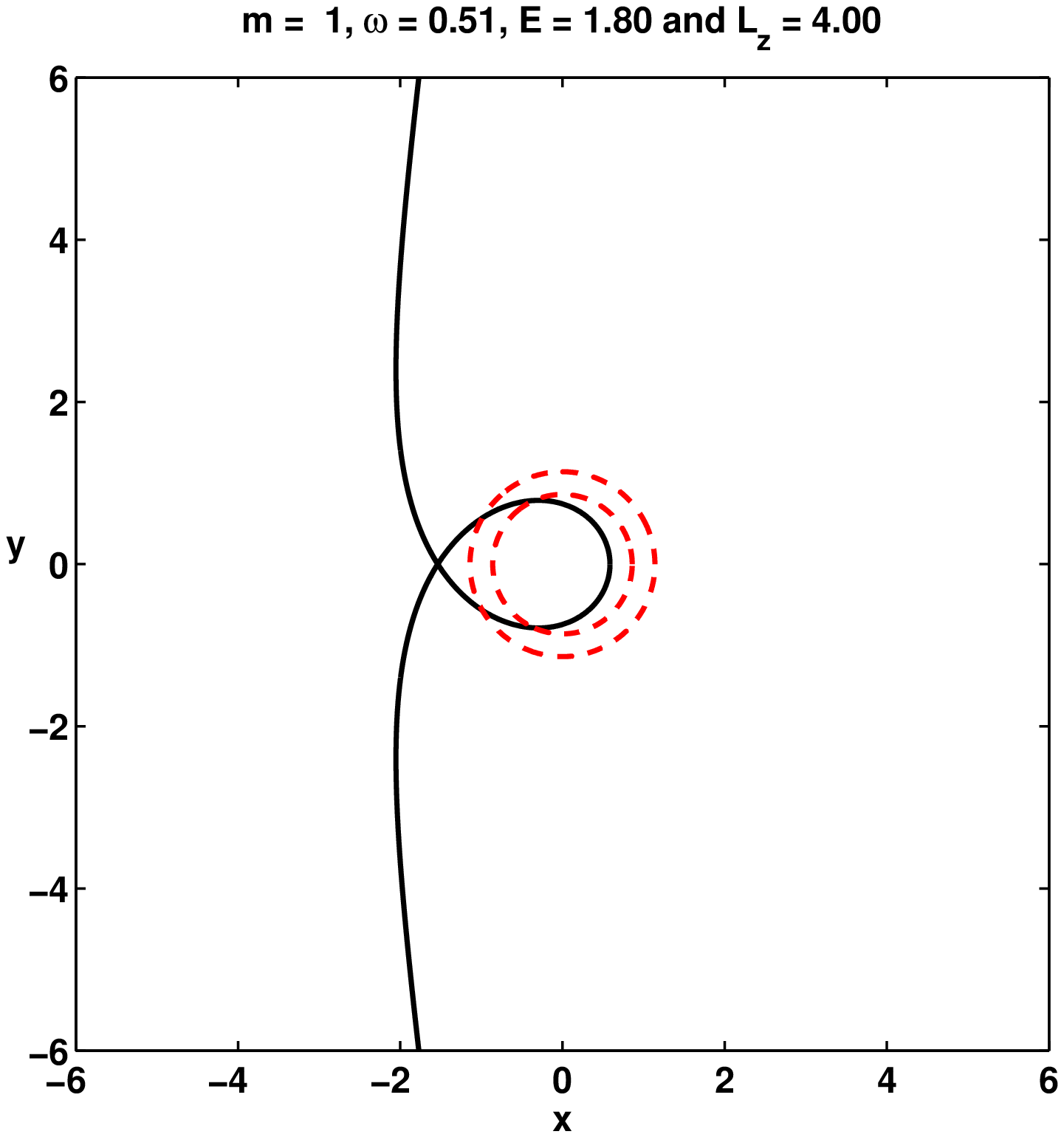}}
    \subfigure[$\omega=1.53$]{\label{orbit2b}\includegraphics[scale=0.30]{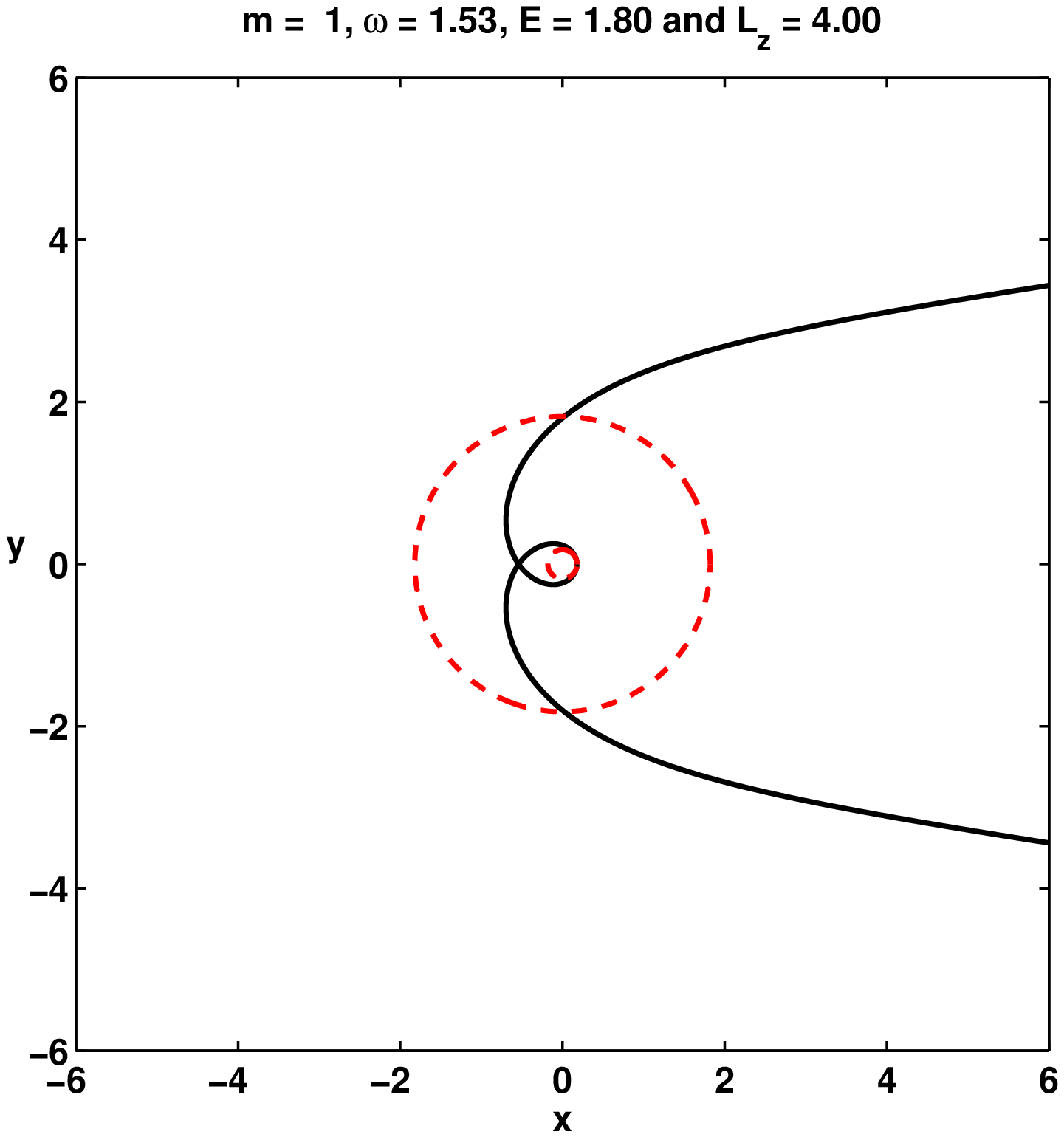}}
    \subfigure[$\omega=50.00$]{\label{orbit2c}\includegraphics[scale=0.30]{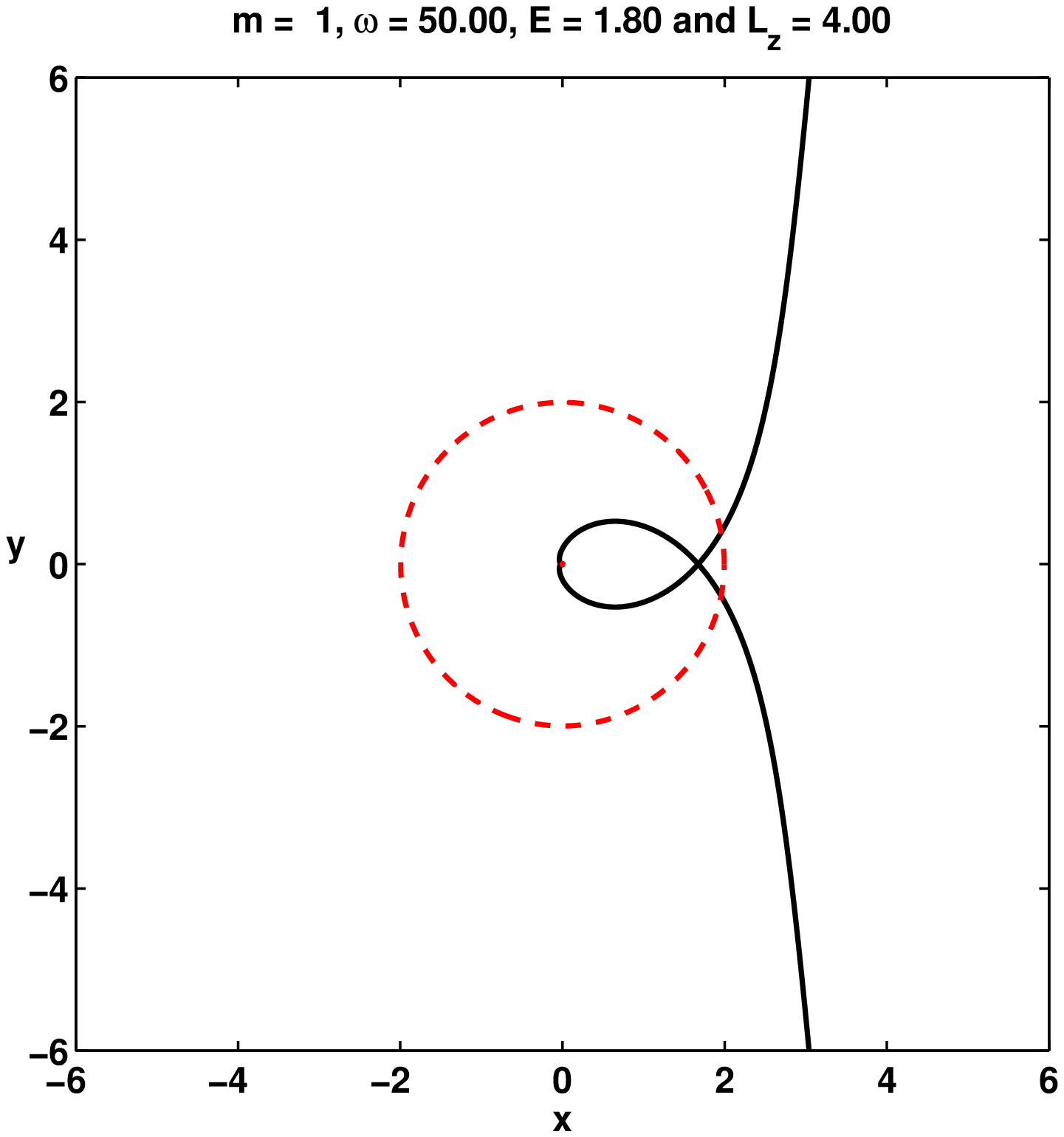}}
    \end{center}
   \caption{Examples of two-world escape orbits (TEO) of a massive test particle  ($\varepsilon = 1$) with $E = 1.8$, $L_z = 4$ 
in the space-time of a KS black hole 
with $m = 1$ and different values of $\omega$. The
dashed circles correspond to the two horizons of the KS space-time.}
\label{escape_massive}  
\end{figure}

These observations lead to the following conclusion for the types of orbits possible which have turning points at the 
minimal radius $r=r_{\rm min}$ close to $r=0$: 
test particles would move on manyworld bound orbits (MBO) with $r_{\rm min} < r < r_{\rm max}$
or on two-world escape orbits (TEO) with  $r_{\rm min} < r \le \infty$ but can never reach $r=0$. 
In comparison to bound orbits (BO) and escape orbits (EO), respectively,
test particles moving on manyworld or two-world orbits cross the two horizons in both directions. That this is
always the case for orbits with $r_{\rm min}$ close to $r=0$ can be seen as follows:
since $V_{\rm eff}(r_{\pm})=-\varepsilon\equiv -1$ and the turning points are given by $E^2-\varepsilon=V_{\rm eff}(r)$, 
the value of $r_{\rm min}$ is always smaller than $r_-$ and the value
of $r_{\rm max}$ is always larger than $r_+$. 
In the Schwarzschild space-time manyworld or two-world orbits are not possible: a particle crossing the horizon would
always end at the physical singularity at $r=0$.
Note that in the KS space-time we also have bound orbits that are comparable to the bound orbits
existing in the Schwarzschild space-time. The two regions in which manyworld bound orbits and bound orbits, respectively, exist 
are shown in Fig.\ref{bound_orbits} for $m=1$ and two
different values of $\omega$. In region 1 we have manyworld bound orbits, while in region
2 we have bound orbits. The effective potential varies only little in the region 2 at large $r$ when changing
$\omega$ from $10^4$ to $0.51$, while in region 1 at small $r$ it varies strongly.  
The above results are summarized in the $(E^2-\varepsilon)$-$(1/L_z^2)$-plot (see Figs. \ref{fig2b}-\ref{fig2c}).

\begin{table}[t]
\begin{center}
\begin{tabular}{cccc}\hline
region & positive zeros & range of $r$ & orbit \\ \hline\hline
I & 4 &  
\begin{pspicture}(-2,-0.2)(3,0.2)
\psline[linewidth=0.5pt]{->}(-2,0)(3,0)
\psline[linewidth=0.5pt,doubleline=true](1.0,-0.2)(1.0,0.2)
\psline[linewidth=0.5pt,doubleline=true](-0.5,-0.2)(-0.5,0.2)
\psline[linewidth=1.2pt]{*-*}(-1.0,0)(1.5,0)
\psline[linewidth=1.2pt]{*-*}(2.0,0)(2.6,0)
\end{pspicture} 
 & MBO, BO  \\ \hline
II & 3 & 
\begin{pspicture}(-2,-0.2)(3,0.2)
\psline[linewidth=0.5pt]{->}(-2,0)(3,0)
\psline[linewidth=0.5pt,doubleline=true](1.0,-0.2)(1.0,0.2)
\psline[linewidth=0.5pt,doubleline=true](-0.5,-0.2)(-0.5,0.2)
\psline[linewidth=1.2pt]{*-*}(-1.0,0)(1.5,0)
\psline[linewidth=1.2pt]{*-}(2.0,0)(3,0)
\end{pspicture} 
 & MBO, EO \\ \hline
III & 2  & 
\begin{pspicture}(-2,-0.2)(3,0.2)
\psline[linewidth=0.5pt]{->}(-2,0)(3,0)
\psline[linewidth=0.5pt,doubleline=true](1.0,-0.2)(1.0,0.2)
\psline[linewidth=0.5pt,doubleline=true](-0.5,-0.2)(-0.5,0.2)
\psline[linewidth=1.2pt]{*-*}(-1.0,0)(1.5,0)
\end{pspicture} 
& MBO \\  \hline
IV & 1  & 
\begin{pspicture}(-2,-0.2)(3,0.2)
\psline[linewidth=0.5pt]{->}(-2,0)(3,0)
\psline[linewidth=0.5pt,doubleline=true](1.0,-0.2)(1.0,0.2)
\psline[linewidth=0.5pt,doubleline=true](-0.5,-0.2)(-0.5,0.2)
\psline[linewidth=1.2pt]{*-}(-1.1,0)(3,0)
\end{pspicture} 
 & TEO \\  \hline \hline
\end{tabular}
\caption{Types of orbits of massive test particles
 in the KS space-time. 
The thick lines represent the range of the orbits. 
The turning points are shown by thick dots. 
The horizons are indicated by double vertical lines. 
\label{table1}}
\end{center}
\end{table}

The shaded region is bounded by two curves, 
the one at larger $E^2-\varepsilon$ representing the maximum of the potential
and the other one the local minimum of the potential at large $r$. The dark shaded region with $E^2-\varepsilon < 0$ (region I)
corresponds to the values of $E^2$ and $L_z^2$ for which ${\cal E}-V_{\rm eff}(r)$ has four positive real-valued zeros.
Hence, there are two different types of orbits: a manyworld bound orbit (MBO) as well as a bound orbit (BO). 
The light shaded region with $E^2 - \varepsilon > 0$ (region II) corresponds to the values of 
$E^2$ and $L_z^2$ for which ${\cal E}-V_{\rm eff}(r)$ has three positive real-valued zeros and hence we 
have a manyworld bound orbit (MBO) as well as
an escape orbit (EO). In the white region with $E^2-\varepsilon < 0$ (region III) ${\cal E}-V_{\rm eff}(r)$ possesses 
two positive, real-valued zeros such that the corresponding orbit is a manyworld bound orbit (MBO).
Finally in the white region with $E^2 -\varepsilon > 0$ (region IV) 
${\cal E}-V_{\rm eff}(r)$ has 
one positive, real-valued
zero and the corresponding orbit is a two-world escape orbit (TEO). These results are also summarized in Table \ref{table1}.

Note that the orbits existing in this space-time are very similar to the ones in the Reissner-Nordstr\"om space-time \cite{chandra,valeria}.
Comparing the case for $\omega=5.1$ with that for $\omega=5.1\cdot 10^4$, we observe that the features of the
plot do not vary much. This is also true for even smaller values of $\omega$. 

Massive test particles with $L_z=0$ move on radial geodesics with $\varphi=const.$.
In this case, the minimum of the effective
potential is at $r=r_0$ such that
\begin{equation}
 r_{0}=\left(\frac{m}{2\omega}\right)^{1/3} \ \ \ {\rm and} \ \ \ V_{\rm eff}(r_0)= -\left(2\omega m^2\right)^{1/3}  \ .
\end{equation}
Note that for $L_z=0$ we can write the effective potential as $V_{\rm eff}(r)=N^2(r)-1$. This leads to the observation
that the value of the effective potential at the horizons $r_{\pm}$ is given by $V_{\rm eff}(r_{\pm})=-1$. Since for
black hole solutions we will always have $V_{\rm eff}(r_0) \le -1$ we find that for massive
particles $r_- \le r_0 \le r_+$. We show the effective potential
for $\varepsilon=1$, $m=1$ and different values of $\omega$ in Fig.\ref{radial_potential}.
We thus find two different possible radial orbits depending on the value of $E^2$. 
For $E^2-1 > 0$ the particle moving on  a radial geodesic will
be able to reach the physical singularity at $r=0$, while that with $ -1 < E^2-1 < 0$ cannot reach $r=0$ and will be deflected
at a finite value of $r=r_{\rm min}$. Moreover, this latter particle cannot reach $r=\infty$ and will be deflected at $r=r_{\rm max}$.
The turning points are at $r_{\rm min,max}$ with
\begin{equation}
 r_{\rm min,max}=\frac{1}{2}\left[\frac{2\omega m \pm \sqrt{4\omega^2m^2 +2\omega(E^2-1)^3}}{\omega (1-E^2)}\right]    \ .
\end{equation}
Since $V_{\rm eff}(r_{\rm min,max})=E^2-1 \ge -1$, we find that $r_{\rm min} \le r_- \le r_0 \le r_+ \le r_{\rm max}$.
The particles are thus trapped on radial manyworld orbits moving from
$r_{\rm max}$ to $r_{\rm min}$ and back to $r_{\max}$ and crossing the horizons in both directions while doing so.

\subsubsection{Massless test particles}

In Figs. \ref{fig1d}-\ref{fig1f} we show how the effective potential $V_{\rm eff}(r)$ for a massless test particle ($\varepsilon=0$)
changes for different values of $L_z$ and $\omega$ with $m=1$. The potential possesses always two 
extrema: one maximum, which for $\omega=\infty$ is located at $r=3m$ and a minimum. The value of this
minimum is negative and increases with decreasing $\omega$ becoming equal to zero in the extremal limit.
The existence of a minimum is a new feature as compared to the Schwarzschild case. Again, we have an infinite potential
barrier at $r=0$. Hence, in contrast to the Schwarzschild case we can now have 
three positive, real-valued zeros of ${\cal E}-V_{\rm eff}(r)$ if $E^2$ is smaller than the maximum of the potential.
The possible orbits are a manyworld bound orbit (MBO) on which the particle crosses both horizons with $r_{\rm min} < r < r_{\rm max}$.
In addition there is an
escape orbit (EO) with $r_{\rm min} < r \le \infty$, where the value of $r_{\rm min}$ fulfills $r_{\rm min} > r_+$. 
These escape orbits are very similar to the ones existing in the Schwarzschild space-time.
For $E^2$ larger than the maximum
of the potential there is only one positive, real-valued zero of ${\cal E}-V_{\rm eff}(r)$  and 
the particle moves on a two-world escape orbit (TEO). The argument that the particle should always cross
both horizons for the manyworld bound orbit and the two-world escape orbit (TEO), respectively, is similar to the massive case: 
since $V_{\rm eff}(r_{\pm})=-\varepsilon\equiv 0$ and the turning points are given
by $E^2=V_{\rm eff}(r)$ we find that $r_{\rm min}$ is always smaller than $r_-$ and $r_{\rm max}$ is always larger than
$r_+$. Again, test particles with non-vanishing angular momentum cannot reach the singularity at $r=0$.  
Our results are summarized in Table \ref{table2}.

\begin{table}[t]
\begin{center}
\begin{tabular}{ccc}\hline
positive zeros & range of $r$ & orbit \\ \hline\hline
3 & 
\begin{pspicture}(-2,-0.2)(3,0.2)
\psline[linewidth=0.5pt]{->}(-2,0)(3,0)
\psline[linewidth=0.5pt,doubleline=true](1.0,-0.2)(1.0,0.2)
\psline[linewidth=0.5pt,doubleline=true](-0.5,-0.2)(-0.5,0.2)
\psline[linewidth=1.2pt]{*-*}(-1.0,0)(1.5,0)
\psline[linewidth=1.2pt]{*-}(2.0,0)(3,0)
\end{pspicture} 
 & MBO, EO \\ \hline
1  & 
\begin{pspicture}(-2,-0.2)(3,0.2)
\psline[linewidth=0.5pt]{->}(-2,0)(3,0)
\psline[linewidth=0.5pt,doubleline=true](1.0,-0.2)(1.0,0.2)
\psline[linewidth=0.5pt,doubleline=true](-0.5,-0.2)(-0.5,0.2)
\psline[linewidth=1.2pt]{*-}(-1.1,0)(3,0)
\end{pspicture} 
 & TEO \\  \hline \hline
\end{tabular}
\caption{Types of orbits of massless test particles
 in the KS space-time. 
The thick lines represent the range of the orbits. 
The turning points are shown by thick dots. 
The horizons are indicated by double vertical lines. 
\label{table2}}
\end{center}
\end{table}

The effective potential for radially moving test particles $(L_z=0)$ is $V_{\rm eff}(r)\equiv 0$.
Hence, all massless test particles will reach the singularity at $r=0$ on radial geodesics.

\begin{figure}[h!]
  \begin{center}
    \subfigure[$\omega = 0.51$]{\label{orbit3a}\includegraphics[scale=0.30]{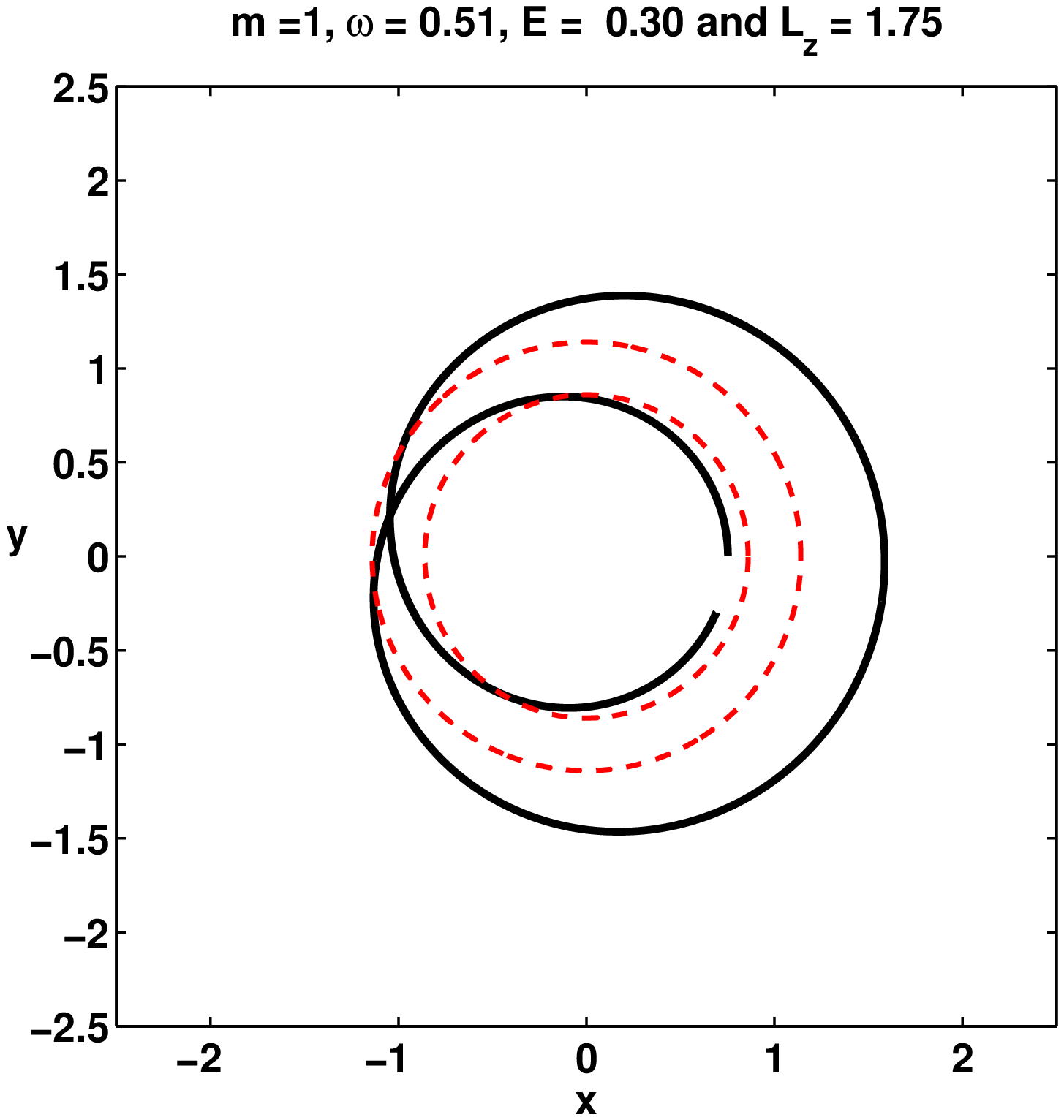}}
    \subfigure[$\omega = 1.53$]{\label{orbit3b}\includegraphics[scale=0.30]{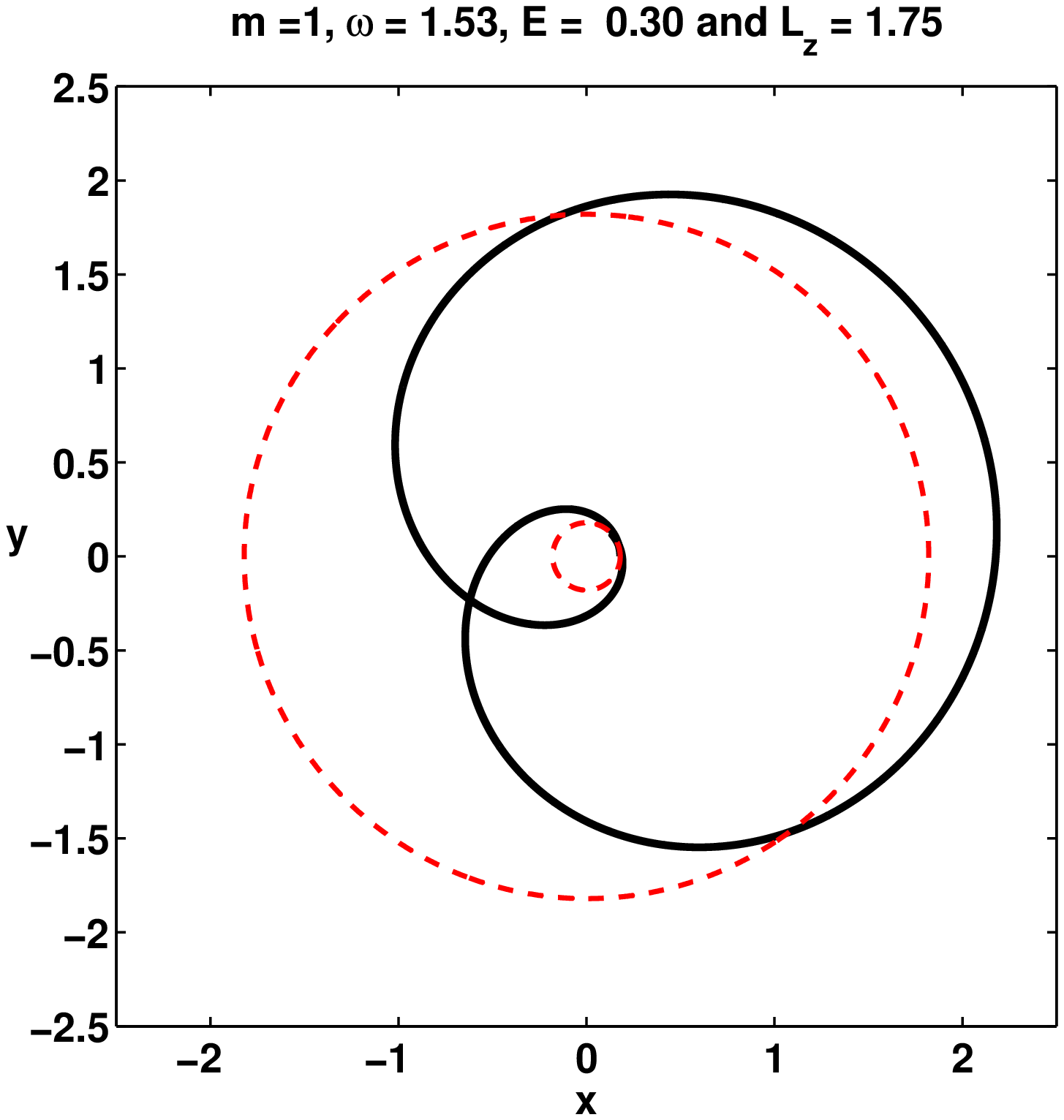}}
    \subfigure[$\omega = 50.00$]{\label{orbit3c}\includegraphics[scale=0.30]{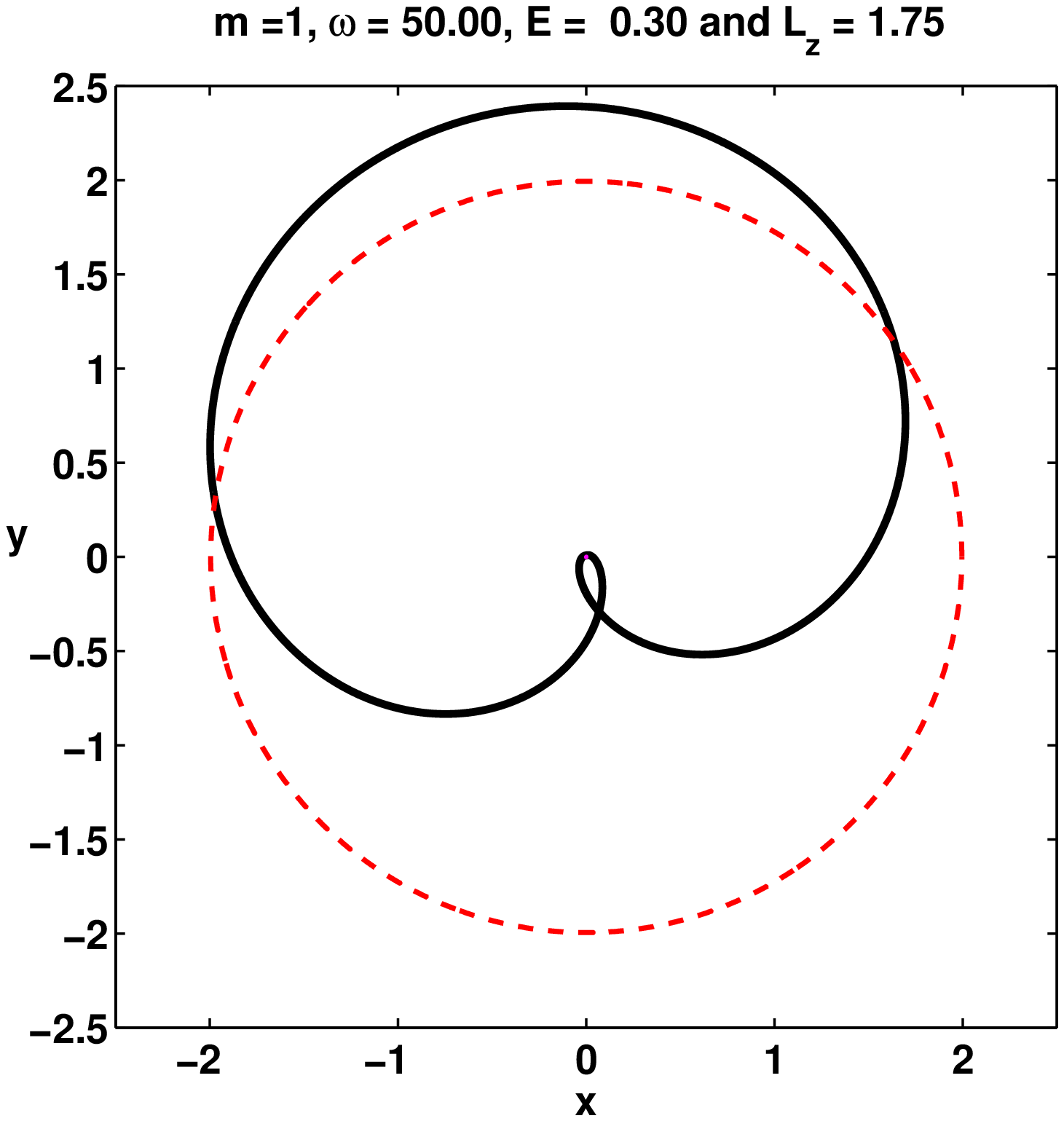}}
    \end{center}
   \caption[Optional caption for list of figures]{
Examples of manyworld bound orbits (MBO) of a massless test particle  ($\varepsilon = 0$) with $E = 0.3$, $L_z = 1.75$ 
in the space-time of a KS black hole 
with $m = 1$ and different values of $\omega$. The
dashed circles correspond to the two horizons of the KS space-time. Note that we are plotting one radial period during which
the particle moves twice from $r_{\rm min}$ to $r_{\rm max}$ and back again.}
\label{bound_massless} 
 \end{figure}

\subsection{Examples of orbits}
In order to find the motion of massive and massless particles in the KS space-time, we have solved the 
equation (\ref{geo1}) numerically using the ODE solver of MATLAB that has a 4th order Runge-Kutta method implemented. 
The relative (resp. absolute) errors of the solution are on the order
of $10^{-12}$ ($10^{-15}$). 

\subsubsection{Massive test particles}
\label{massive_orbits}

In Fig.\ref{fig3} we show manyworld bound orbits (MBO) and bound orbits (BO)
(region 1 and 2, see Fig.\ref{bound_orbits}), respectively, for $E=0.99$ and $L_z=7.0$. 
In region 2 the test particle moves on a nearly circular orbit
with radius much larger than the horizon radii. The shape of the orbit varies only little when changing $\omega$. 
In region 1, on the other hand, the orbit is quite different for $\omega=10^4$ as compared to $\omega=0.51$. For both
values of $\omega$, the test particle crosses the two horizons in both directions suggesting that these bound orbits are 
manyworld bound orbits (MBO). Note that
this is similar to the case of test particles moving in the Reissner-Nordstr\"om space-time \cite{chandra,valeria}.

Due to the infinite potential barrier at $r=0$ a test particle with 
non-vanishing angular momentum coming from infinity would be reflected at a finite
value of $r$ and would not be able to reach $r=0$ in the KS space-time. This is shown 
in Fig.\ref{escape_massive}, where we give examples of two-world escape orbits (TEO) of a massive test particle
with angular momentum $L_z=4$ and energy $E=1.8$ for different values of $\omega$ and $m=1$. 
For all values of $\omega$, the particle crosses both horizons, but does not reach $r=0$, i.e. the particle
approaches the KS black hole from an asymptotically flat region, crosses both horizons twice and moves away into
another asymptotically flat region.

\subsubsection{Massless test particles}
As stated above we now have the possibility of manyworld bound orbits (MBO) for massless test particles which are not possible in the
space-time of a Schwarzschild black hole.
In Fig.\ref{bound_massless} we give examples of manyworld bound orbits (MBO) of massless test particles with angular momentum
$L_z=1.75$ and energy $E=0.3$. The qualitative features of the orbits are very similar to the
massive case. For all values of $\omega$ the particle crosses both horizons, but due to the infinite potential
barrier can never reach the physical singularity at $r=0$. Note that bound orbits (BO)
of massless test particles moving solely outside the black hole do not exist.

\begin{figure}[h!]
  \begin{center}
    \subfigure[$\omega=0.51$]{\label{orbit4a}\includegraphics[scale=0.30]{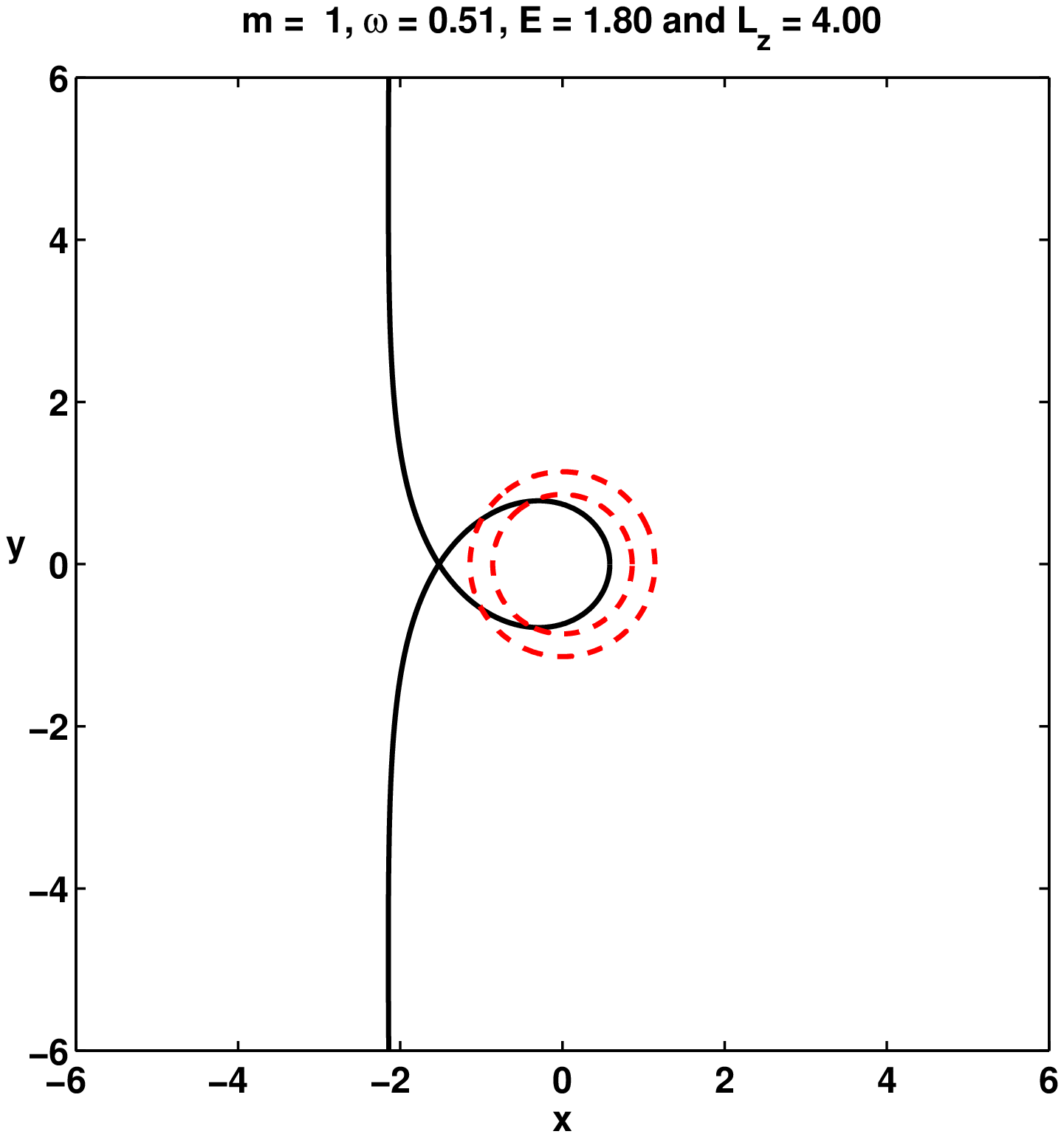}}
    \subfigure[$\omega=1.53$]{\label{orbit4b}\includegraphics[scale=0.30]{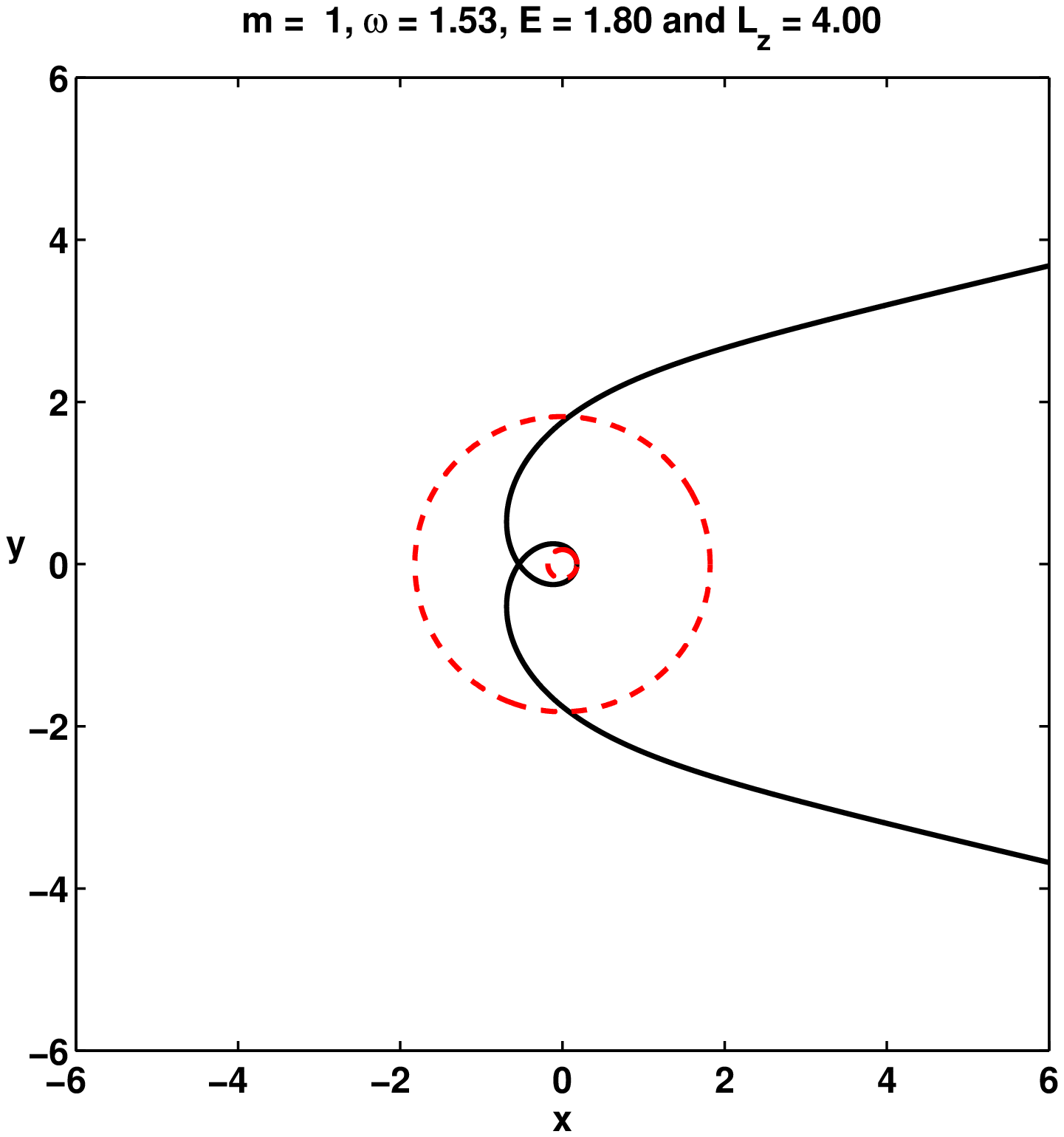}}
    \subfigure[$\omega=50.00$]{\label{orbit4c}\includegraphics[scale=0.30]{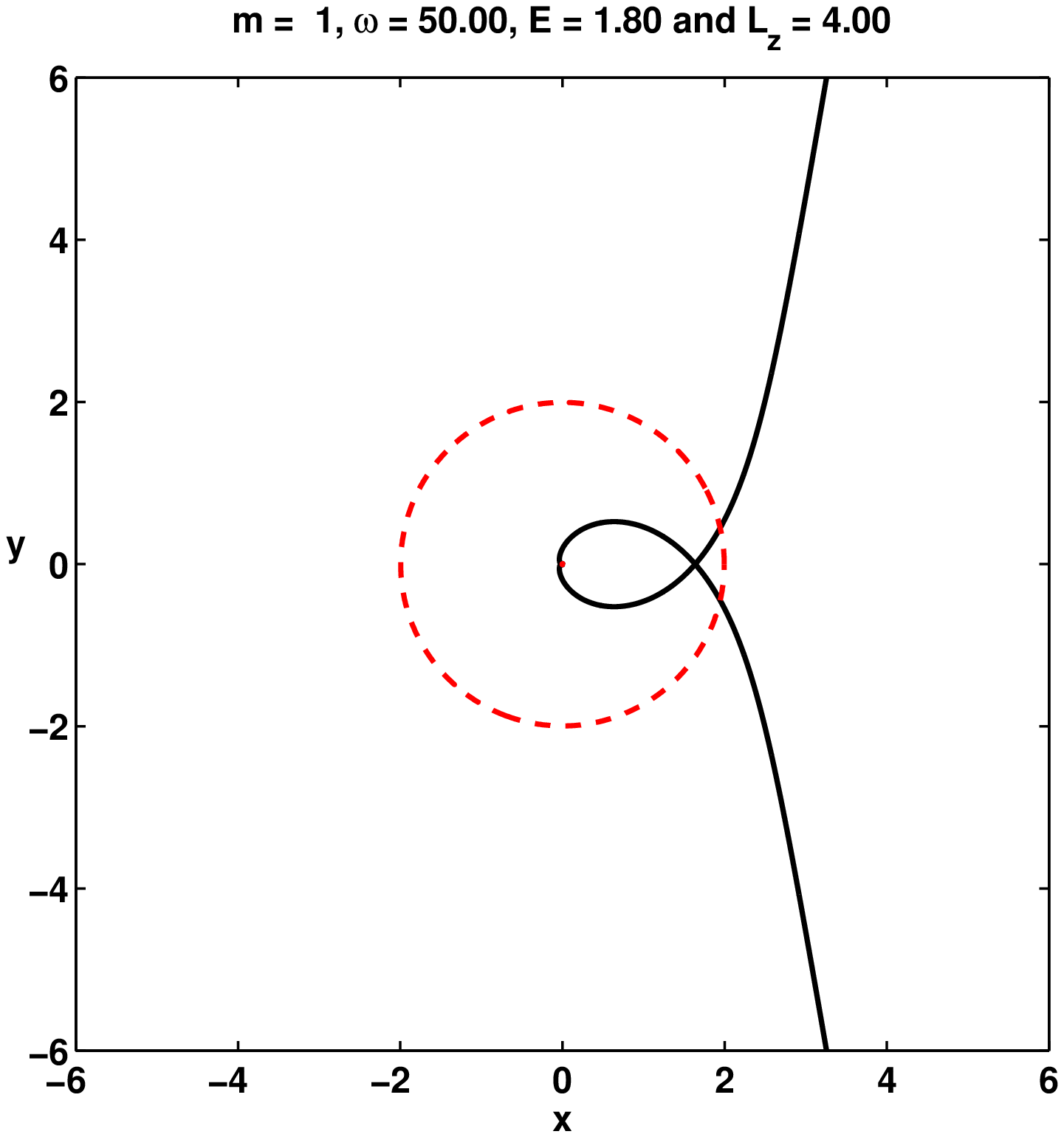}}
    \end{center}
   \caption{Examples of two-world escape orbits (TEO) of a massless test particle  ($\varepsilon = 0$) with $E = 1.8$, $L_z = 4$ 
in the space-time of a KS black hole 
with $m = 1$ and different values of $\omega$. The
dashed circles correspond to the two event horizons of the KS space-time.}
\label{escape_massless1}
  \end{figure}

In Fig.\ref{escape_massless1} we give examples of two-world escape orbits (TEO) of massless test particles with angular
momentum $L_z=4$ and energy $E=1.8$. In this case, the test particle encircles the space-time singularity at $r=0$ and crosses
the horizons while doing so. 

\begin{figure}[h!]
  \begin{center}
    \subfigure[$V_{\rm eff}(r)$]{\label{veff5}\includegraphics[scale=0.30]{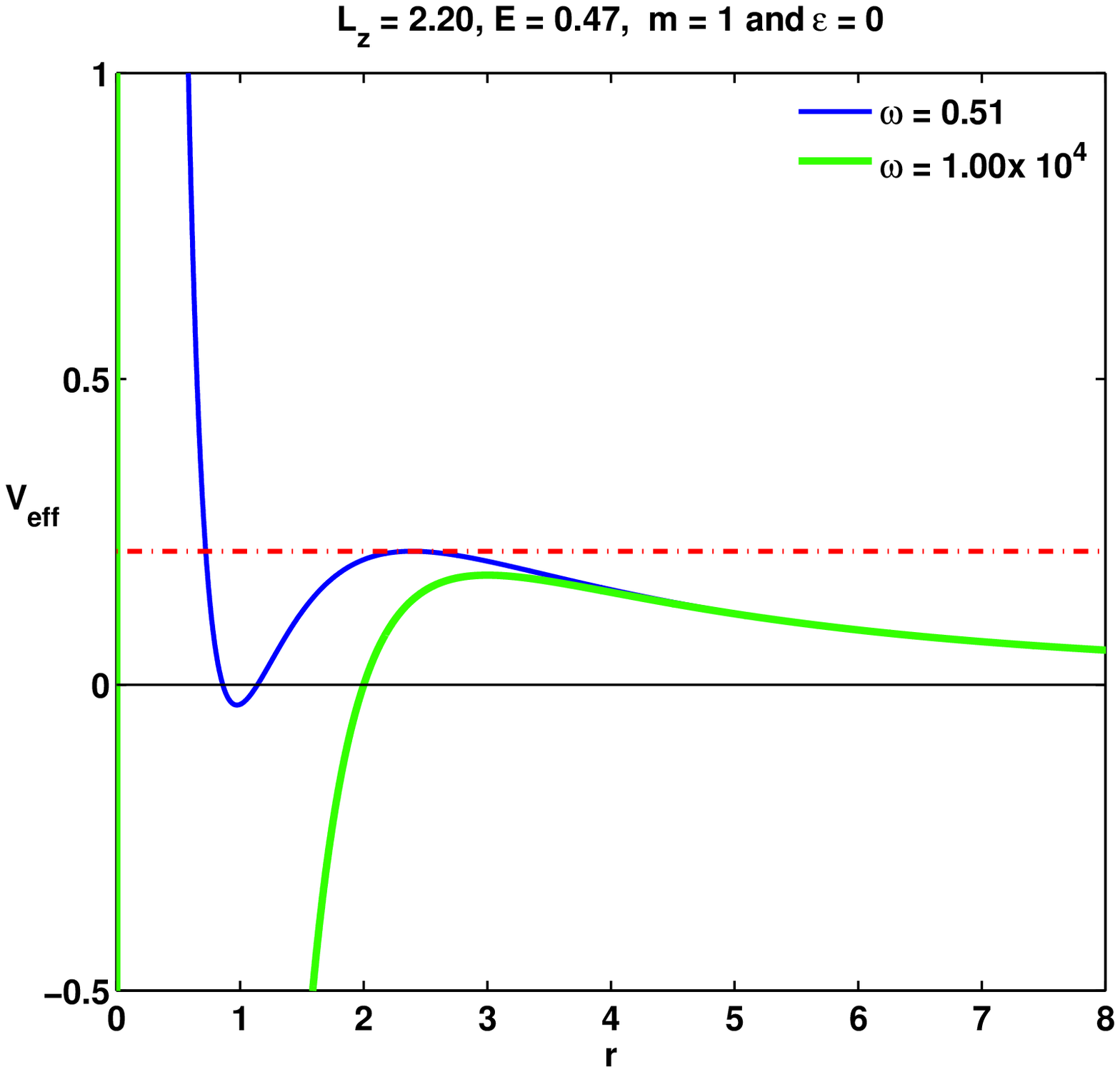}}
    \subfigure[EO]{\label{orbit5a}\includegraphics[scale=0.30]{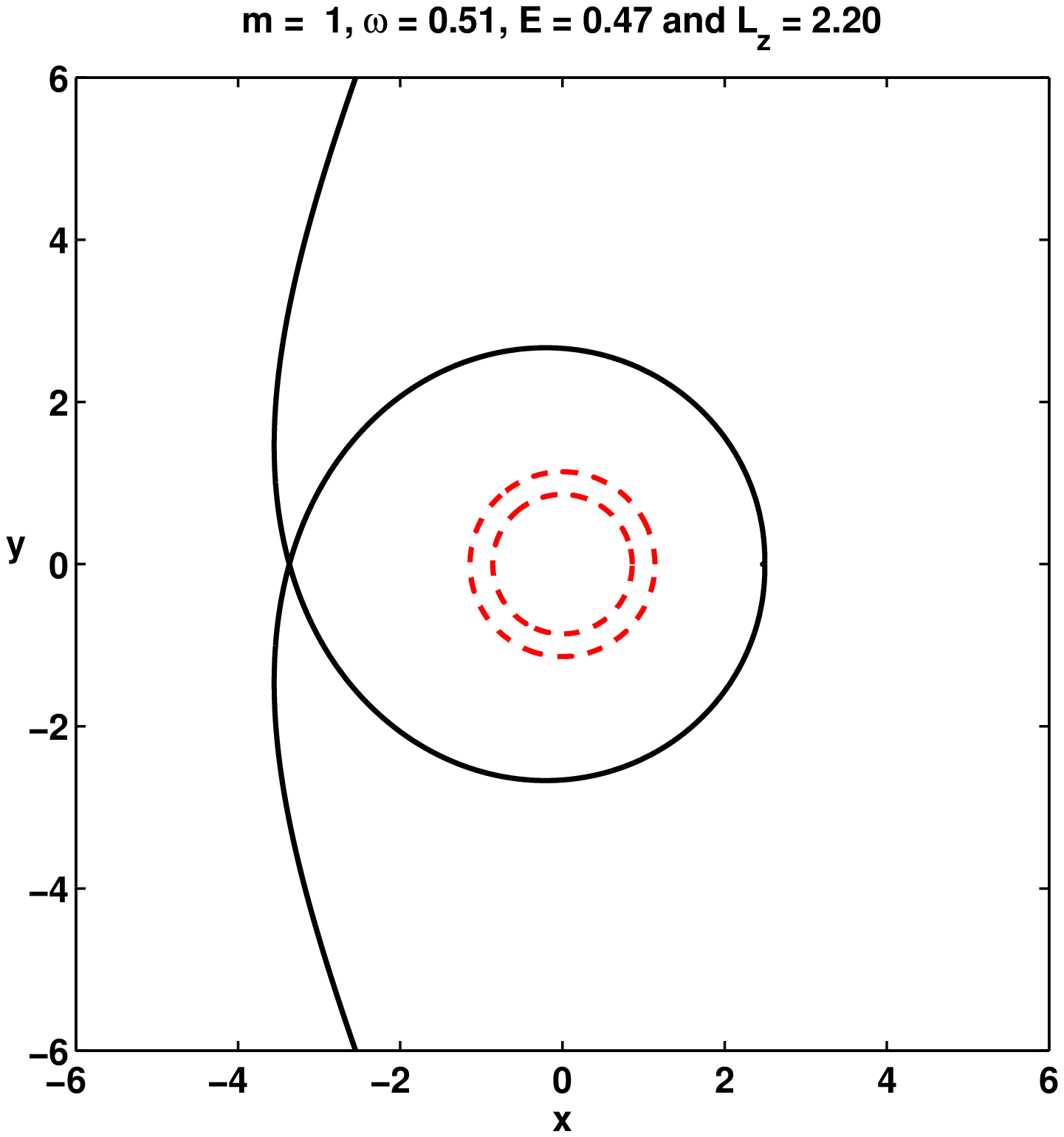}}
    \subfigure[TEO]{\label{orbit5b}\includegraphics[scale=0.30]{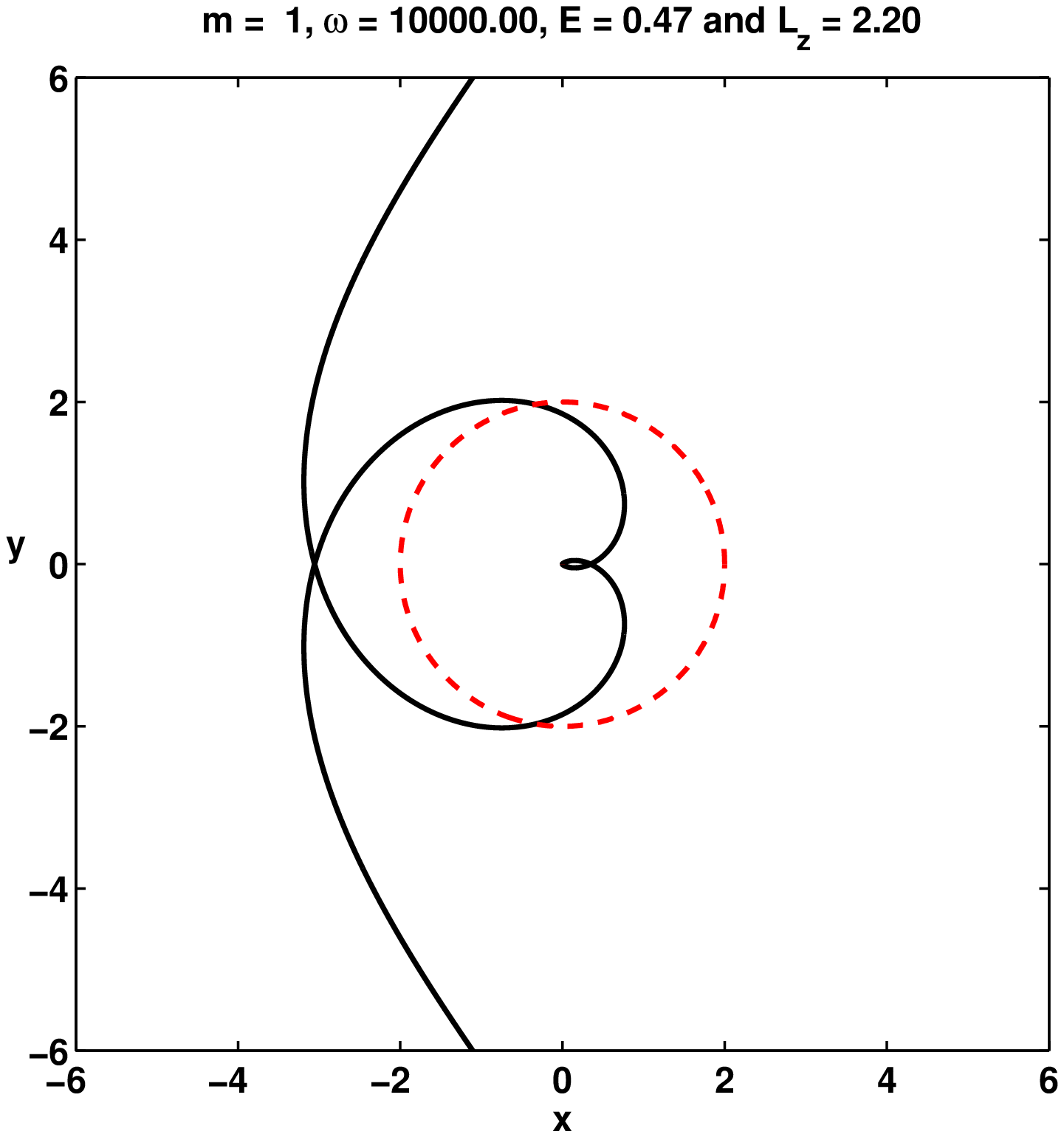}}
    \end{center}
   \caption{Examples of escape orbits (EO) of a massless test particle  ($\varepsilon = 0$) with $E = 0.47$, $L_z = 2.2$ 
that passes very close by a KS black hole  
with $m = 1$ and $\omega=0.51$ (Fig. \ref{orbit5a}). For much larger values of $\omega$ (here: $\omega=10^4$) the particle
crosses the horizon on a two-world escape orbit (TEO) (Fig. \ref{orbit5b}). We also show the corresponding effective potential
(Fig. \ref{veff5}). The red dashed line in Fig. \ref{veff5} corresponds to the value of $E^2$, while the
dashed circles in Fig. \ref{orbit5a}-\ref{orbit5b} correspond to the two event horizons of the KS space-time.}
\label{escape_massless2}
  \end{figure}

In Fig.\ref{escape_massless2} we give an example of an escape orbit (EO) of a massless test particle with
angular momentum $L_z=2.2$ and energy $E=0.47$  that
is deflected by the KS black hole and comes very close to the horizons, but never crosses them. This
is for $m=1$ and $\omega=0.51$ (see Fig.\ref{orbit5a}). For the same values of energy and angular momentum but much larger
values of $\omega$ the test particle
would cross the horizons and move on a two-world escape orbit (TEO) (see Fig.\ref{orbit5b}). 
These orbits should be compared to predictions recently made for massless test particles passing
close by a Kerr black hole \cite{nature}.

\subsection{Observables}

\subsubsection{Perihelion shift}

\begin{figure}[h!]
  \begin{center}
    \subfigure[MBO]{\label{ps1}\includegraphics[scale=0.45]{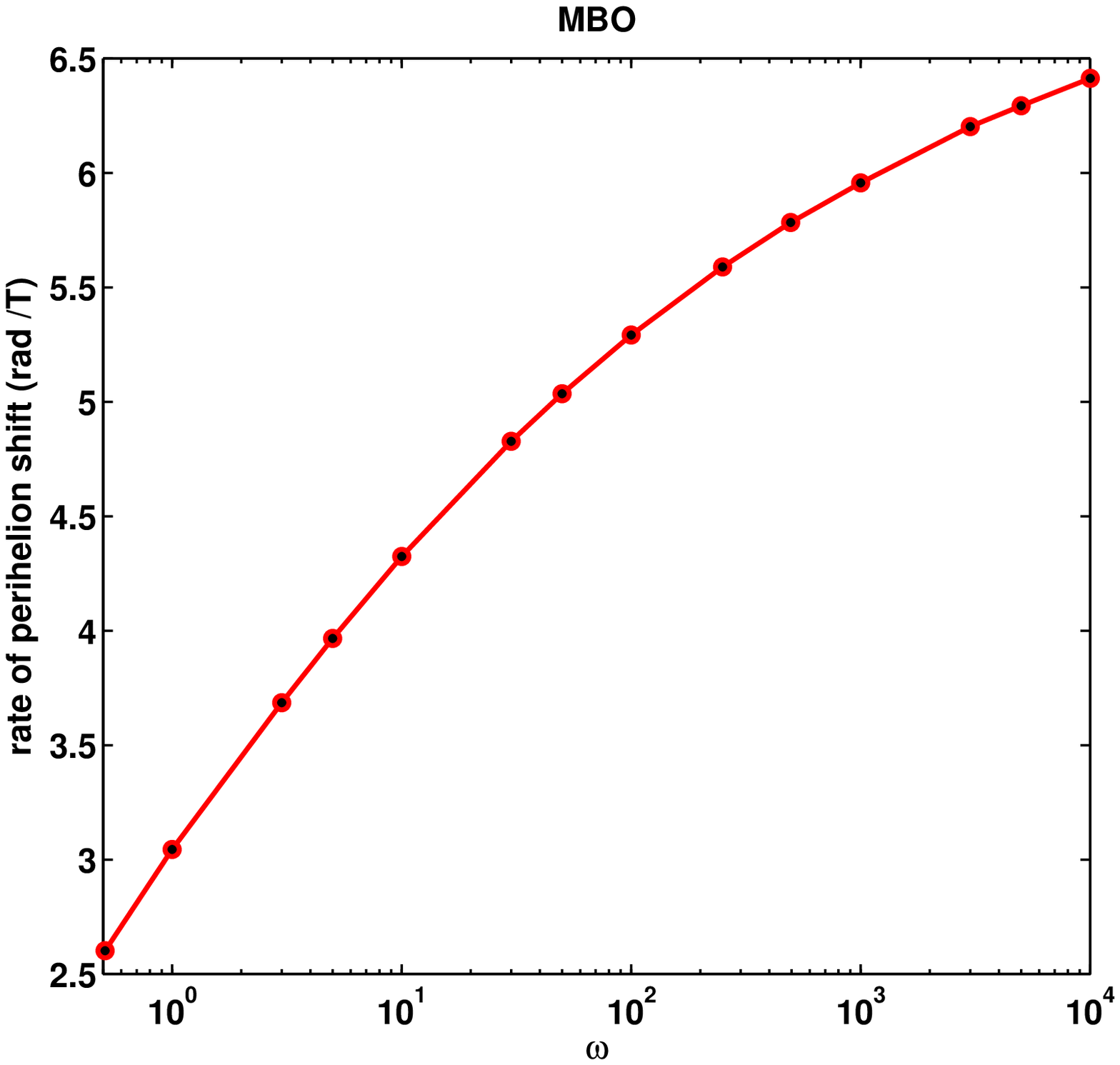}}
    \subfigure[BO]{\label{ps2}\includegraphics[scale=0.45]{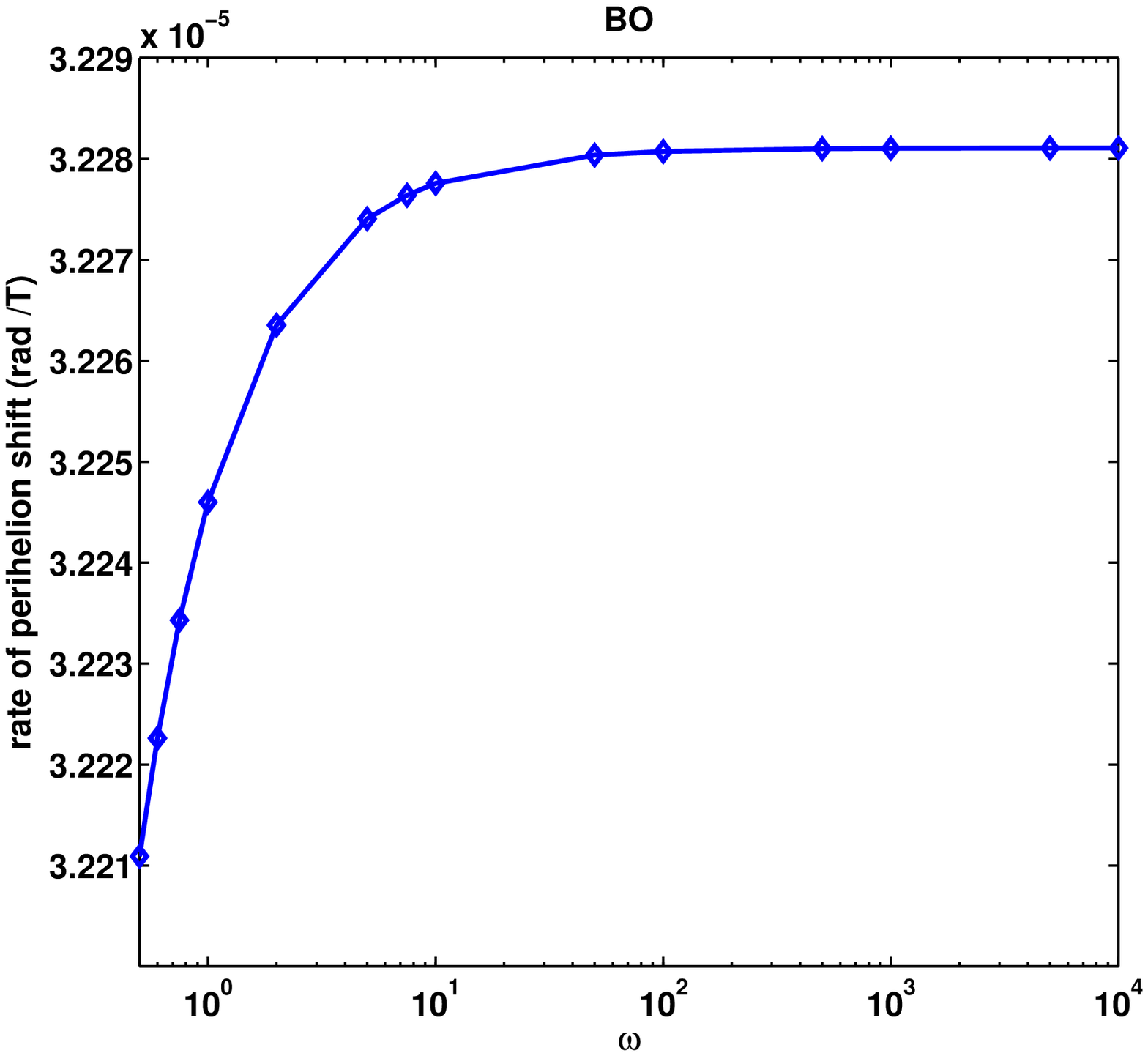}}
    \end{center}
   \caption{The value of the perihelion shift per period $T$ as a function of $\omega$ 
for a massive test particle ($\varepsilon = 1$) with energy  $E^2=0.9787$ and angular momentum $L_z = 7.00$ in the space-time of a
KS black hole with $m =1$. We show the perihelion shift per period $T$ for a manyworld bound orbit (MBO) (left) and
for a bound orbit (BO) (right), respectively.
\label{ps}}
  \end{figure}

The perihelion shift of a bound orbit of a massive test particle in the space-time of a KS black hole can be calculated by using
(\ref{eq2}). We find for the perihelion shift $\delta\varphi$ and the period $T$ of the motion of a massive test particle 
from $r_{\rm min}$ to $r_{\rm max}$ and back again
\begin{equation}
 \delta \varphi = 2\int\limits_{r_{\rm min}}^{r_{\rm max}}   
\frac{L_z dr}{r^2\sqrt{E^2 - f\left(1+L_z^2/r^2\right)}} - 2\pi  \ , \  T=2\int\limits_{r_{\rm min}}^{r_{\rm max}}   
\frac{dr}{\sqrt{E^2 - f\left(1+L_z^2/r^2\right)}}   \ .
\end{equation}
Our results for $m=1$ are shown in Fig.\ref{ps}, where we give the value of the rate of the perihelion shift $\delta\varphi/T$ in
dependence on $\omega$. In Fig.\ref{ps1} we show the perihelion shift for a manyworld bound orbit (MBO), while in Fig.\ref{ps2}
we show that of a bound orbit (BO). 
We observe that the perihelion shift of the manyworld bound orbit (MBO) is much larger than that of the bound orbit (BO). For both
types of orbits the perihelion shift increases with increasing $\omega$.

\begin{figure}[h!]
  \begin{center}
    \subfigure[$\varepsilon=1$]{\label{diff1}\includegraphics[scale=0.45]{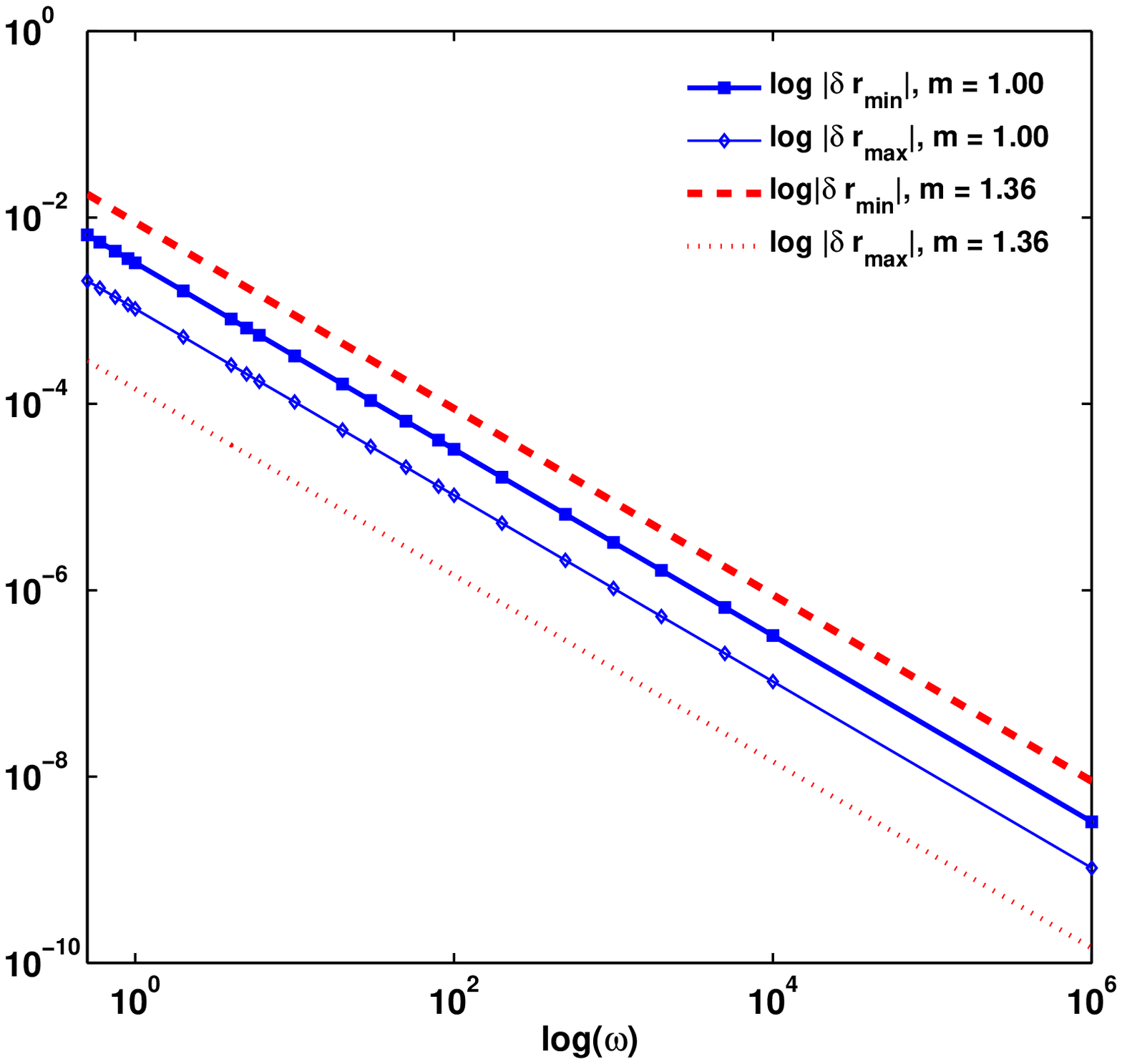}}
    \subfigure[$\varepsilon=0$]{\label{diff2}\includegraphics[scale=0.45]{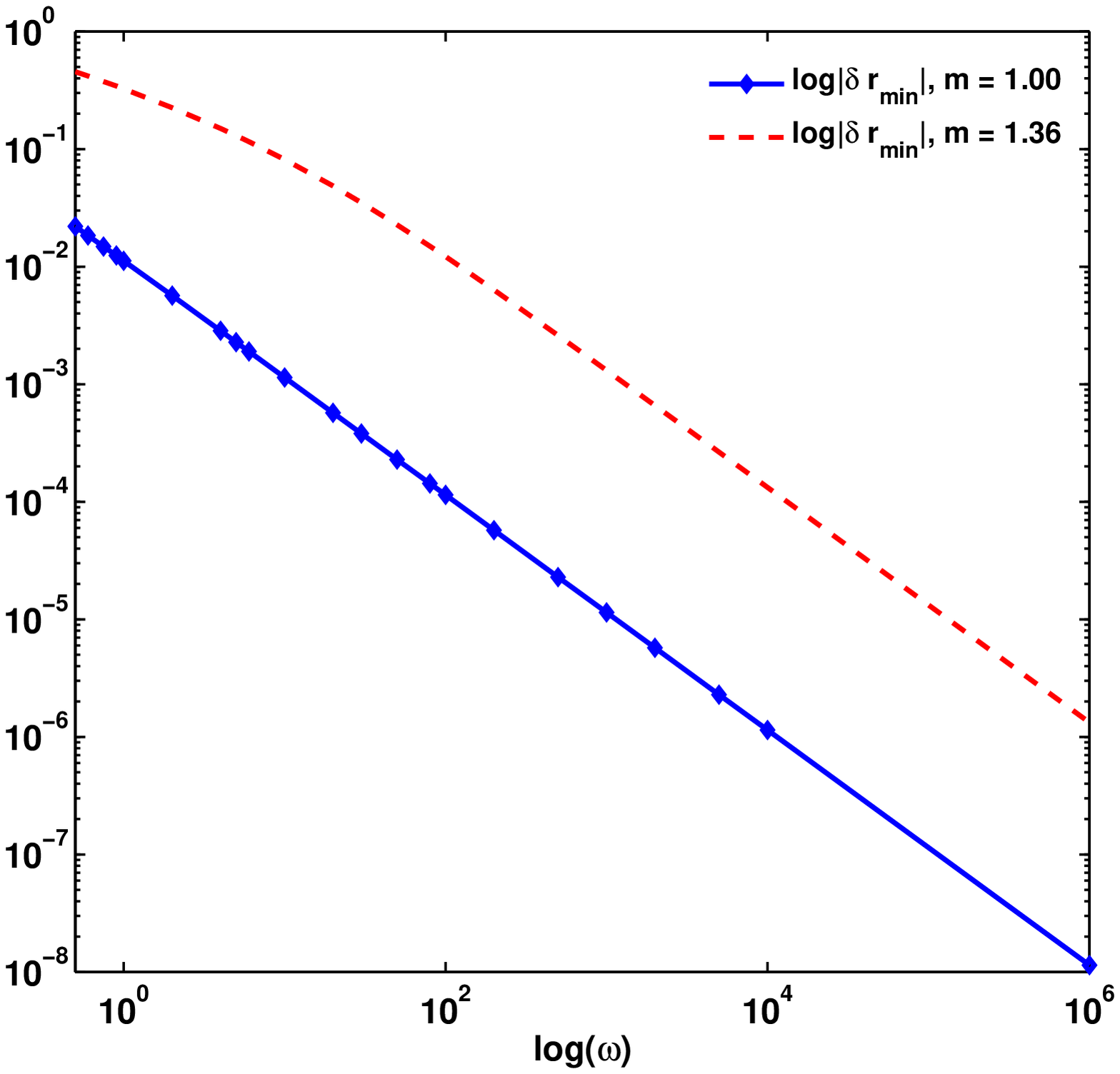}}
    \end{center}
   \caption{The absolute value of the difference $\delta r_{\rm min}=r_{\rm min,KS}-r_{\rm min,S}$ ($\delta r_{\rm max}=r_{\rm max,KS}-r_{\rm max,S}$) 
between the minimal (maximal) radius of the bound orbit (BO) of a massive particle in the KS space-time $r_{\rm min,KS}$ ($r_{\rm max,KS}$) 
and the minimal (maximal) 
radius of a bound orbit (BO) in the Schwarzschild space-time $r_{\rm min,S}$ ($r_{\rm max,S}$) is shown in 
dependence on $\omega$ for two different values of $m$ (left).
We also give the difference $\delta r_{\rm min}$ for an escape orbit (EO) of a massless test particle (right).
In both cases the energy of the particle is $E=0.99$ and the angular momentum $L_z = 7.00$.}
  \end{figure}

We can compare the perihelion shift of a bound orbit (BO) in the KS black hole space-time with that in the Schwarzschild space-time.
Note that the bound orbits of test particles with energy  $E^2=0.9787$ and angular momentum $L_z = 7.00$ are nearly circular 
(see Fig.\ref{fig3c} and Fig.\ref{fig3d}). Hence, it is a good approximation to use the perturbative formula for the Schwarzschild
space-time which gives the perihelion shift as function of the mass of the central object $M_{\rm S}$ 
\cite{chandra}
\begin{equation}
 \left(\delta\varphi\right)_{\rm S}=6\pi \frac{m_{\rm S}^2 c^2}{\tilde{l}^2}  \ \ , \ \ m_{\rm S}=\frac{GM_{\rm S}}{c^2} \ \  ,  \ \
\frac{1}{\tilde{l}}=\frac{1}{2}\left(\frac{1}{r_{\rm max}} + \frac{1}{r_{\rm min}}\right) \ \ ,
\end{equation}
where $G$ is Newton's constant and $c$ is the speed of light. We have then computed the value of the perihelion shift $\delta\varphi$
of the bound orbit  of a test particle with energy $E^2=0.9787$ and angular momentum $L_z = 7.00$ in the KS space-time with $m=1$ 
and several values of $\omega$. Setting this value equal to $(\delta\varphi)_{\rm S}$ we can find the corresponding
mass $m_{\rm S}$ that is necessary to obtain the same value of the perihelion shift in the Schwarzschild space-time. 
We find that $m_{\rm S}\approx 1.109$ when comparing with the KS space-time for values of $\omega$ between unity and $10^4$, i.e.
$m_{\rm S}$ does not vary much. This leads to the following observation~: to have the same perihelion shift in the KS space-time as compared to the
Schwarzschild space-time we need a smaller mass of the central body. Moreover, the value of $r_{\rm min}$ 
(respectively $r_{\rm max}$) is larger (smaller)
for a bound orbit (BO) in the KS space-time as compared to a bound orbit (BO) in the Schwarzschild space-time with the same
value of the perihelion shift. This would be a way to distinguish KS from Schwarzschild space-times. This is shown in Fig.\ref{diff1}, where
we give the difference $\delta r_{\rm min}=r_{\rm min,KS} - r_{\rm min,S}$ of the minimal radius
in the KS space-time $r_{\rm min,KS}$ and in the Schwarzschild space-time $r_{\rm min,S}$ as function of $\omega$ for
two different values of $m$. 
We also give the
value of $\delta r_{\rm max}= r_{\rm max,KS} - r_{\rm max,S}$ of the difference of the maximal radius
in the KS space-time $r_{\rm max,KS}$ and the maximal radius in the Schwarzschild space-time $r_{\rm max,S}$.  
Note that the value of the radius of the black hole is between $m$ in the extremal
limit and $2m$ in the Schwarzschild limit. For a stellar black hole with radius $10km$, this would correspond to
masses between $3.39$ solar masses (for $\omega=\infty$) and $6.78$ solar masses (for the extremal limit). We observe that the difference
decreases with increasing $\omega$ (as expected). For increasing $m$ both the difference $\delta r_{\rm min}$ as well as
the difference $\delta r_{\rm max}$ increase. For this note that $\delta r_{\rm max}$ is in 
fact negative and we are giving the absolute value here such that the absolute value of $\delta r_{\rm max}$ decreases with increasing $m$.



\subsubsection{Light deflection}

\begin{figure}[h!]
  \begin{center}
    \subfigure[TEO]{\label{ld1}\includegraphics[scale=0.45]{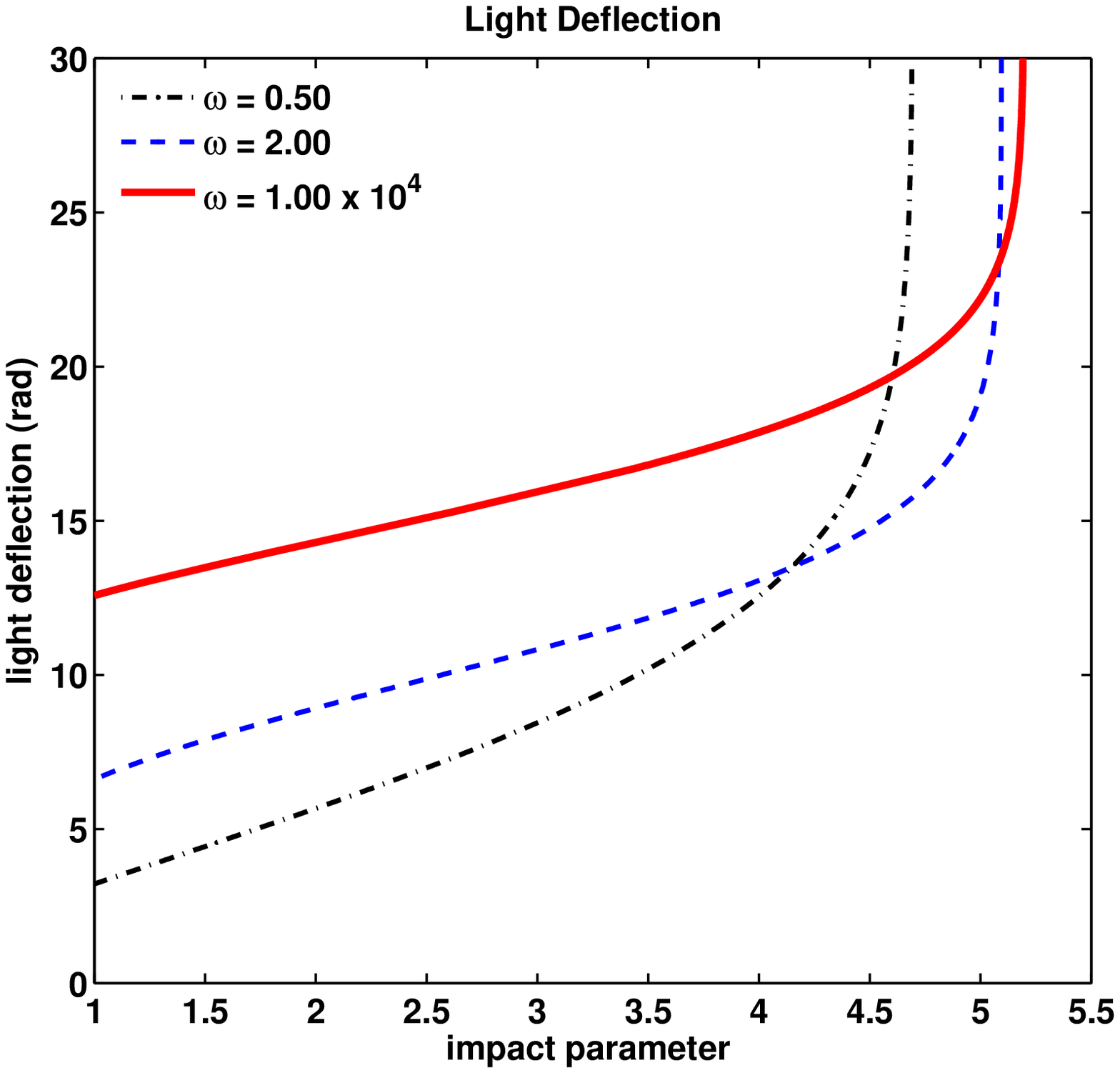}}
    \subfigure[EO]{\label{ld2}\includegraphics[scale=0.45]{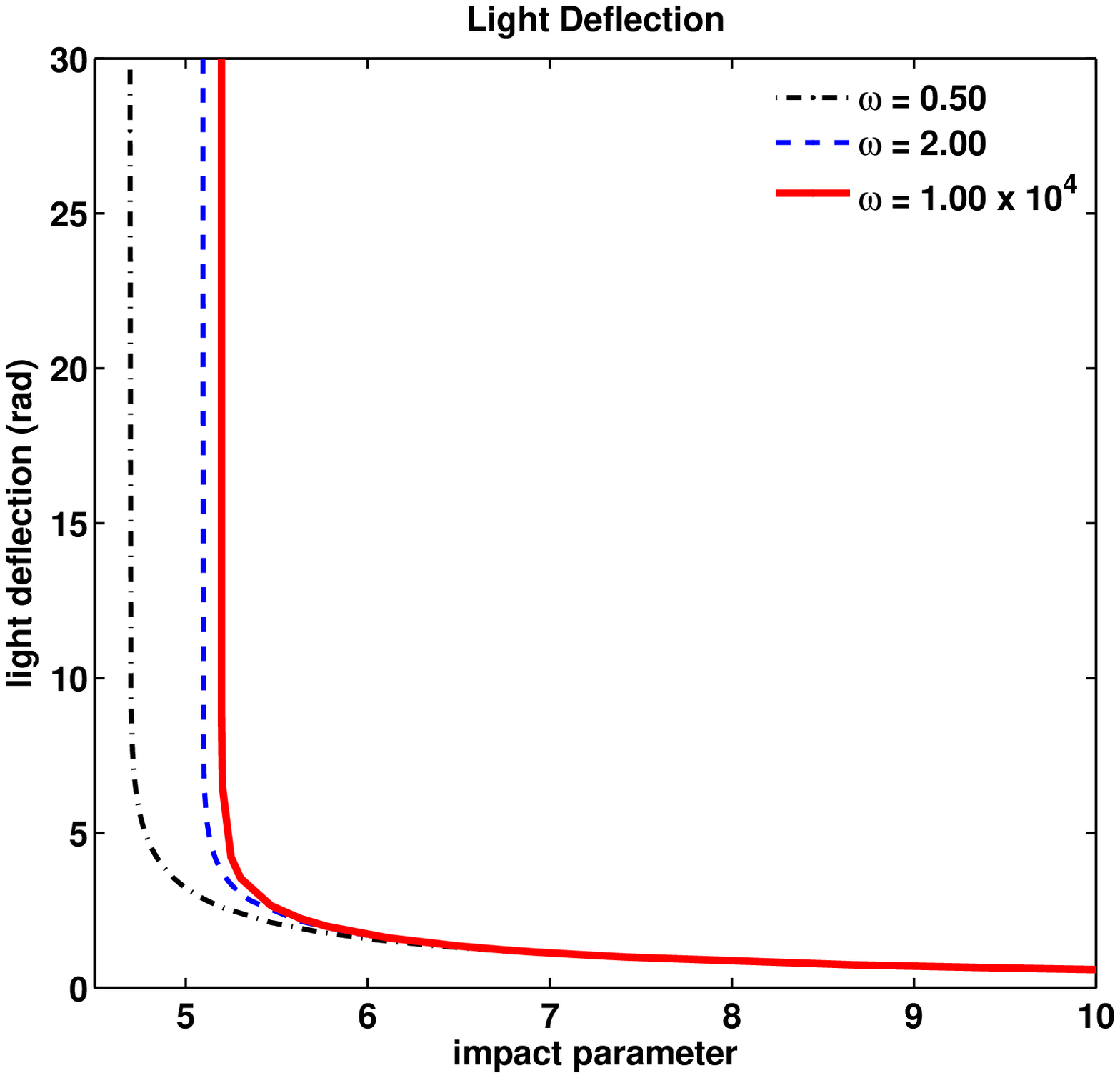}}
    \end{center}
   \caption{The value of the light deflection of a massive test particle in the space-time of the KS black hole with $m= 1$ 
as a function of 
the impact parameter $b=L_z/E$. We give the light deflection for the two-world escape orbit (TEO) (left) and for
the escape (EO) (right) for three different values of $\omega$. Note that the light deflection
diverges at $b=b_{\rm crit}$ with $b_{\rm crit}=4.6937$ for $\omega=0.50$, $b_{\rm crit}=5.0950$ for $\omega=2.00$
and $b_{\rm crit}=5.1961\approx \sqrt{27}$ for $\omega=10^4$, respectively.
\label{ld}}
  \end{figure}

The deflection of light by a KS black hole can be calculated by using
 (\ref{eq2})  for an escape orbit of a massless test
particle ($\varepsilon=0$). The light deflection then reads
\begin{equation}
 \widetilde{\delta \varphi} = 2\int\limits_{r_{\rm min}}^{\infty}   
 \frac{L_z dr}{r^2\sqrt{E^2 - f L_z^2/r^2 }}  - \pi  \ ,
\end{equation}
where $r_{\rm min}$ is the minimal radius of the orbit.
Our results for $m=1$ are shown in Fig.\ref{ld}, where we give the value of the light deflection in
dependence on the impact parameter $b$, which is equal to $L_z/E$ in the case that the initial value of $r$ is equal
to infinity. Note that for values of $\widetilde{\delta \varphi}$
larger than $2\pi$ the massless test particle first encircles the black hole once or several times 
before going back to infinity. In Fig.\ref{ld1} we show the light deflection of the two-world escape orbit (TEO), while
in Fig.\ref{ld2} we show the light deflection of the escape orbit (EO).

For the latter case, we observe that the light deflection increases with decreasing impact parameter. This is
very similar to the Schwarzschild space-time. Lowering the value of the impact parameter further we find that the light deflection
diverges at a critical value $b=b_{\rm crit}$. This critical value depends on $\omega$ and decreases with decreasing
$\omega$: at $\omega=10^4$ the value is close to the Schwarzschild value $b_{\rm crit}=5.1961\approx \sqrt{27}$, while 
$b_{\rm crit}=5.0950$ for $\omega=2.00$ and $b_{\rm crit}=4.6937$ for $\omega=0.50$. Lowering the impact parameter
even further we find that the light deflection now decreases with decreasing impact parameter. This is a 
new feature as compared to the Schwarzschild space-time, which however also exists in the Reissner-Nordstr\"om case \cite{chandra,valeria}.
This phenomenon can be explained when considering the form
of the effective potential. Lowering the impact parameter $b$ is comparable to fixing $L_z$ and increasing the energy $E$.
For small value of $E$ (large values of $b$) there are three positive real zeros of ${\cal E}-V_{\rm eff}(r)$
and the corresponding orbits are a manyworld bound orbit (MBO) and an escape orbit (EO). Increasing $E$ (decreasing $b$)
we would then find a critical value of $E$ for which $E^2$ is equal to the value of the maximum of the effective potential.
This corresponds to an unstable circular orbit for which the value of the light deflection diverges. 
Increasing $E$ (lowering $b$) further, $E^2$ has only one intersection point at positive $r$ with the effective potential and this
corresponds to a two-world escape orbit (TEO).

We also find that the light deflection for the escape orbit (EO) decreases with decreasing $\omega$, while for the
two-world escape orbits the dependence on $\omega$ depends on the value of the impact parameter. For very small
impact parameter, the light deflection decreases with decreasing $\omega$, while for $b$ close to $b_{\rm crit}$ 
it increases with decreasing $\omega$. 

We can again compare with the Schwarzschild case. For large impact parameter $b$ the light deflection in the Schwarzschild space-time
can be approximated by \cite{chandra}
\begin{equation}
 \widetilde{\delta \varphi}_{\rm S}=\frac{4m_{\rm S}}{b} \ .
\end{equation}
We have then computed $\widetilde{\delta \varphi}$ for test particles with $L_z=5.1961$ and different impact parameters
$b=L_z/E$  in the KS space-time and set these values equal to $\widetilde{\delta \varphi}_{\rm S}$ to find the
corresponding values $m_{\rm S}$. We find that for $\omega=2$, $m=1$ we need to choose $m_{\rm S}\approx 1.168$ for impact
parameter $b=40$ and $m_{\rm S}\approx 1.170$ for impact parameter $b=29.85$, respectively, to get the same value of
the light deflection. For increasing $\omega$ the corresponding $m_{\rm S}$ decreases, e.g. $m_{\rm S}\approx 1.15$ for
$\omega=10^4$. The conclusion is very similar to the one in the case of the perihelion shift: to find the same
value of the light deflection in the KS space-time as compared to the Schwarzschild space-time the mass of the
central body has to be smaller. $r_{\rm min}$ of the escape orbit (EO) in the KS space-time is larger as compared to an escape
orbit in the Schwarzschild space-time for the same value of the light deflection. This would be another method to distinguish the KS space-time
from the Schwarzschild space-time and is shown in Fig.\ref{diff2}, where we give the difference $\delta r_{\rm min}=r_{\rm min,KS} - r_{\rm min,S}$
for the escape orbit (EO) of a massless test particle in dependence on $\omega$. Again, we observe that $\delta r_{\rm min}$ decreases
with increasing $\omega$ and increases with increasing $m$.

\section{Conclusions}
In this paper we have studied the motion of massless and massive test particle in the space-time of the
KS black hole, which is a static, spherically symmetric vacuum solution of Ho\v{r}ava-Lifshitz (HL) gravity.
We have taken the viewpoint that Ho\v{r}ava-Lifshitz gravity is essentially a short-distance modification
of General Relativity (GR) and have used the GR geodesic equation. We observe that there are some
new features as compared to the static, spherically symmetric vacuum solution of GR, the Schwarzschild solution.
For massive test particles we find that next to bound orbits there exist manyworld bound orbits on which the test particles
cross the two horizons in both directions.
For massless test particles we can also have manyworld bound orbits, which do not exist in the Schwarzschild case.
There exist also escape orbits, which are comparable to the ones in the Schwarzschild case as well
as two-world escape orbits, which are a new feature.
Due to an infinite angular momentum barrier, test particles with non-vanishing angular momentum 
can never reach $r=0$ - in contrast to the
Schwarzschild case where particles that have crossed the event horizon unavoidably move to $r=0$.
Massless test particles moving on radial geodesics will always go to $r=0$, while massive test particles moving
on these geodesics 
are either trapped on a manyworld radial geodesic if their energy $E < 1$ or they will reach the singularity at $r=0$ for
$E > 1$.

We have also computed the perihelion shift and the light deflection. The rate of the perihelion shift of the manyworld bound orbit
is much larger than that of the bound orbit and for both orbits this rate decreases with decreasing $\omega$, i.e. it
is largest for both types of orbits in the Schwarzschild limit. The light deflection increases
with decreasing impact parameter for escape orbits, but decreases with decreasing impact parameter for 
two-world escape orbits. For escape orbits the light deflection is decreasing for decreasing $\omega$, while
for two-world escape orbits it decreases (resp. increases) for small (large) impact parameter.
Approximate methods have been used in several other papers to constrain the value of 
the parameter $\omega m^2$ \cite{bobo,iorio1,iorio2,liu_lu_yu_lu}. Since we 
believe that constraints from orbits can only be obtained for large value of $\omega$, i.e.
close to the Schwarzschild limit we have not attempted to recompute 
the constraints since we believe that our exact techniques would more or less
give the same numbers as those found in \cite{bobo,iorio1,iorio2,liu_lu_yu_lu}.
The aim of this paper has been to solve the geodesic equation exactly and present 
the complete set of solutions to the
geodesic equation.

Recently, the geodesic equation in another Ho\v{r}ava-Lifshitz black hole space-time
has been solved analytically in terms of hyperelliptic functions \cite{ehkkls}. It seems possible that
in some limiting cases of the KS black hole space-time considered here, we can also find 
analytic solutions. This is currently under investigation \cite{ehkkls_new}.
\\
\\
{\bf Acknowledgments}
The work of PS has been supported by DFG grant HA-4426/5-1. VK has been supported by the DFG.
CL thanks the Center of Excellence QUEST for support. 

\providecommand{\bysame}{\leavevmode\hbox to3em{\hrulefill}\thinspace}

\end{document}